\documentclass[twocolumn]{article}

\usepackage[a4paper,top=2cm,bottom=2cm,left=1cm,right=1cm]{geometry}
\usepackage{abstract}
\usepackage{amsmath,amsfonts}
\usepackage{algorithmic}
\usepackage{algorithm}
\usepackage{array}
\usepackage{authblk}
\usepackage{cite}
\usepackage{enumitem}
\usepackage{graphicx}
\usepackage{hyperref}
\usepackage{multicol}
\usepackage{stfloats}
\usepackage{tabu}
\usepackage{textcomp}
\usepackage{url}
\usepackage{verbatim}
\usepackage[caption=false,font=normalsize,labelfont=sf,textfont=sf]{subfig}
\begin{document}

\title{To Measure What Isn't There --- Visual Exploration of Missingness Structures Using Quality Metrics}

\author[1]{Sara Johansson Fernstad}
\author[1]{Sarah Alsufyani}
\author[2,3,4]{Silvia Del Din}
\author[2,3,4]{Alison Yarnall} 
\author[2,3,4]{Lynn Rochester}

\affil[1]{School of Computing, Newcastle University, UK.}
\affil[2]{Translational and Clinical Research Institute, Newcastle University, UK.} 
\affil[3]{NIHR Newcastle Biomedical Research Centre, Newcastle University, UK.}
\affil[4]{The Newcastle upon Tyne Hospitals NHS Foundation Trust, UK.}

\maketitle

\begin{abstract}
  \textbf{This paper contributes a set of quality metrics for identification and visual analysis of structured missingness in high-dimensional data. Missing values in data are a frequent challenge in most data generating domains and may cause a range of analysis issues. Structural missingness in data may indicate issues in data collection and pre-processing, but may also highlight important data characteristics. While research into statistical methods for dealing with missing data are mainly focusing on replacing missing values with plausible estimated values, visualization has great potential to support a more in-depth understanding of missingness structures in data. Nonetheless, while the interest in missing data visualization has increased in the last decade, it is still a relatively overlooked research topic with a comparably small number of publications, few of which address scalability issues. Efficient visual analysis approaches are needed to enable exploration of missingness structures in large and high-dimensional data, and to support informed decision-making in context of potential data quality issues. This paper suggests a set of quality metrics for identification of patterns of interest for understanding of structural missingness in data. These quality metrics can be used as guidance in visual analysis, as demonstrated through a use case exploring structural missingness in data from a real-life walking monitoring study. All supplemental materials for this paper are available at \url{https://doi.org/10.25405/data.ncl.c.7741829}.}
\end{abstract}


\section{Introduction}

Datasets with missing values, also known as incomplete or missing data, are a common challenge in data analysis that may often cause issues such as biased, uncertain and unreliable results. A large range of statistical methods have been developed for dealing with missing values, which are mainly focused on replacing missing data with plausible values (known as imputation) \cite{Carpenter2013}. The appropriateness and analytical impact of such methods are often dependent on the structures and distribution of missing values in the data. 

Visualization can facilitate increased awareness of the existence of missing values and of structures relating to these, and highlight potential issues and data uncertainties that have arisen during data generation and pre-processing. Furthermore, visualization of missing data can reveal important information related to the missingness, such as patients missing medical appointments where the missingness itself may indicate a health issue. Missing data visualization can also support decision-making as to how the missing values can be most appropriately dealt with. The application of suitable statistical methods for imputation requires understanding of the patterns of missingness, and questions such as if records are missing at random or if there are structured patterns needs to be taken into consideration. 

The challenge and importance of understanding missingess structures in data has been increasingly recognised in visualization literature in the last decade, including the design of novel visualization approaches, e.g. \cite{valero2019plot, fernstad2021explore, jimenez2022graphical}, and studies evaluating the effect of missing data visualization on data interpretation, confidence, and decision making, e.g., \cite{Andreasson2014, Fernstad2019, song2018, song2021understanding, bauerle2022did}. Nonetheless, little effort has been made to address scalability issues in context of missing data visualization. 
The high-dimensionality of data in combination with complex missingness structures has however been highlighted as a particular challenge of proposed methods \cite{fernstad2021explore}. 

Quality Metrics (QM) are measures utilised to capture data properties that are useful for extracting meaningful information in data \cite{Bertini2011}. QM are often used to overcome scalability issues by guiding the user to views of interest, or by mitigating cognitive overload by filtering less meaningful views and data \cite{behrisch2018}. In context of missingness structures, Johansson Fernstad \cite{Fernstad2019}, and Johansson Fernstad and Glen \cite{Fernstad2014} suggested a set of structures of particular relevance for analysis of missing data, namely \textit{Amount Missing}, \textit{Joint Missingness}, and \textit{Conditional Missingness}. This paper builds on their work and contributes by: 
\begin{itemize}
    \item defining a set of QM that captures relevant aspects of these missingness structures in incomplete data;
    \item providing multiple examples of how QM as a concept can be flexibly utilised in visual analysis to reveal patterns in high-dimensional data, through attribute layout and selection as well as visual representation of QM values;
    \item contextualizing the complexities of missing data analysis and demonstrating the utility of the proposed QM through a use case from a real-life walking monitoring study.
\end{itemize}

The paper is structured as follows: Section \ref{sec:background} provides an overview of related work. Section \ref{sec:QM} presents a set of QM for identification of missingness structures, and section \ref{sec:visualinvestigation} demonstrates how these metrics can be flexibly used for visual investigation of structural missingness. Section \ref{Sec:ICICLE} presents a use case exemplar of the application of the QM for exploration of missingness structures in a real-life mobility monitoring dataset, followed by discussion and conclusions in section \ref{sec:conclusions}.

\section{Background}\label{sec:background}

\subsection{Analysis of Incomplete Data}
Missing values can be caused by a range of reasons. For example, survey respondents may avoid answering particular questions; participants in longitudinal studies may not take part in all steps, leading to missing values for certain time points; subjects may drop out or be excluded from clinical trials; and data capturing devices, such as IoT sensors, may fail. When dealing with incomplete data it is usually unknown to the analyst what value the data would have taken if it had been observed. At best a reasonably good estimation of a plausible value can be obtained. 

Analysis of data with a large amount of missing values is a complex problem where the missingness and the method for dealing with it may cause bias, uncertainty and reliability issues. Different to many other uncertainty problems, the missing data issues cannot normally be overcome by increased sample sizes, since the number of missing values often increase alongside the data size. The effect missing values have on analysis results depends both on the missingness mechanism, described in more detail in section \ref{section::misspatterns}, and on how the missing values are handled. It may be greatly affected by the degree of missingness as well as the distribution of missing values across the dataset. 

The two main approaches to dealing with missing values are removal and imputation. Removal is when data points or items with missing values are removed prior to analysis. This approach carries a considerable risk of heavily biased results, unless the missing values are missing completely at random. Imputation on the other hand is when missing values are replaced by estimated values. A large number of imputation methods are available, ranging from replacement with arithmetic mean or random draws from representative distributions, to complex multiple imputation methods that combine several imputations following a set of rules \cite{Carpenter2013}. Imputed values may bias and affect the analysis results, depending on the appropriateness of the imputation method and on the distribution of missing and recorded values.

\subsection{Visualization of Missing Values}
Knowledge about the absence of data values and of structures relating to this absence can in itself sometimes be more informative than estimated values, as exemplified by Fielding et al.\ \cite{Fielding2009} and Djurcilov and Pang \cite{Djurcilov2000}. Considering missing values as information bearing signals can provide valuable information, and visualization of missing values can facilitate understanding of missingness structures and data characteristics, as well as highlight potential issues in data gathering, pre-processing and analysis processes. 

The acceleration of data collection over the last two decades has lead to an increased interest in data quality, including incomplete data, with the majority of research into visualization of missing data being published in the last 10-15 years, as shown in Alsufyani et al.'s State-of-the-Art review of missing data visualization \cite{Alsufyani2024star}. This includes a range of studies that evaluate the effect of missing data visualization on data interpretation, confidence, and decision-making, e.g., \cite{Andreasson2014, Fernstad2019, song2018, song2021understanding, bauerle2022did}, as well as the design of novel visualization approaches to support understanding, exploration and decision making in context of missing data, e.g., \cite{valero2019plot, fernstad2021explore, jimenez2022graphical, tierney2023expanding, alsufyani2024vis}. 

While the increase in publications addressing missing data visualization clearly indicates that the research community is recognising the potential of visualization as a tool for understanding missingness structures in data, there are still challenges to be addressed \cite{Alsufyani2024star, fernstad2021explore}. This includes increased generalizability and realism in evaluations, as well as data related challenges such as data heterogeneity and scalability. In terms of scalability, Johansson Fernstad and Johansson Westberg \cite{fernstad2021explore} highlighted the combination of high-dimensionality and complex missingness structures as a particular challenge.

\subsection{Missingness Structures}\label{section::misspatterns}
The missingness mechanism \cite{Rubin1976} is a commonly used model of how the probability of an observation being missing depends on its own value and on the values of other variables. The {\it Missing Completely at Random} (MCAR) mechanism is when the missing values depends neither on the observed nor the missing part of the data; {\it Missing at Random} (MAR) is when the probability of missing values depends on the observed data; and {\it Missing Not at Random} (MNAR) is when the probability of missing values are depending on something that is not recorded. Missingness mechanisms are rarely known prior to analysis, they are fairly complex and may be difficult to apply to an exploratory analysis approach. More recent research suggests patterns that may be more straightforward for describing structural missingness in data. 

Wang and Wang \cite{Wang2007} suggested three patterns, focussing on the distribution of missing values in context of classification data: 1) {\it Missing at Random}, when missing values are randomly distributed; 2) {\it Uneven Symmetric Missing}, when values are missing more often in some variables and may be correlated; and 3) {\it Uneven Asymmetric Missing}, when values are missing unevenly and may be biased towards a particular class. In a later paper \cite{Wang2009} they also suggested four concepts of relevance for understanding the potential impact of missing values on analysis results: 1) {\it Reliability}, which is a ratio between missing and recorded; 2) {\it Hiding}, which aims to reveal the likeliness of an item with a value in a certain range having a missing value in another variable; 3) {\it Complementing}, which is the likeliness of multiple variables having missing values at the same time; and 4) {\it Conditional Effects}, which is the effect missing values may have on the understanding of the problem. 

Based on previous research and interviews with data science practitioners, Johansson Fernstad \cite{Fernstad2019} defined a set of three missingness patterns of relevance for analysing missingness in data -- {\it Amount Missing}, {\it Joint Missingness} and {\it Conditional Missingness}, described in Section \ref{sec:QM} -- which bring together the main characteristics of previous categorizations of missingness patterns, and provide a more straightforward description of structures than, e.g., the missingness mechanism. The research presented in this paper is focussed on methods for visual exploration of missingness structures based on these concepts.

\subsection{Quality Metrics for High Dimensional Data} 
High dimensionality in visualization can be defined as when it becomes challenging to visually extract meaningful relations among dimensions \cite{Bertini2011}. Common visualization methods for multivariate data, such as Parallel Coordinates (PC) \cite{Inselberg1985} and Scatter Plot Matrix \cite{Becker1987}, are useful for datasets with moderately high-dimensionality, but their usability quickly decrease with increasing dimensionality. Extensive overview of visualization systems and methods for analysis of high-dimensional data are available in, e.g., Bertini et al.\ \cite{Bertini2011}, Johansson Fernstad et al.\ \cite{Fernstad2013} and Liu et al.\ \cite{Liu2017}. 

QM are often utilized for tasks such as projection, ordering, abstraction and view optimization \cite{Bertini2011, behrisch2018}. Bertini et al.\ \cite{Bertini2011} define QM as calculated metrics that capture data properties which are useful for the extraction of meaningful information about data. In context of high-dimensional data visualization, a QM can be thought of as a measure of how interesting a dimension, a subset of dimensions or a dimension ordering is, or how well it represents the underlying data. As such it can help the data analyst to concentrate on the most interesting part of the data. The definition of what is interesting is domain and task dependent, as for example demonstrated by Johansson Fernstad et al.\ \cite{johansson2020quality} who defined a set of QM focussing on tasks of relevance for microbial ecology studies, 
and in many cases multiple measures may be relevant \cite{Johansson2009}. 

Behrisch et al.\ \cite{behrisch2018} provide an extensive review and categorisation of the use of QM in visualization, exemplifying the wide range of tasks, patterns and visualizations that have been the focus of QM research. Several examples exist where QM are used to aid interactive visual exploration of dimension spaces, including the work by Johansson Fernstad et al.\ \cite{Fernstad2011, Fernstad2013} where dimensions were represented in context of multiple QM, using a PC that was also used for interactive subset selection. Their approach were in spirit related to methods presented by Turkay et al.\ \cite{Turkay2011, Turkay2012} and Krause et al.\ \cite{Krause2016}, who both link representations of dimension space and item space through QM. Wang et al.\cite{Wang2019} provided subspace comparison through dimension aggregation and incremental analysis. Lehmann et al.\ \cite{Lehmann2015} identified a set of metrics that work similar to human perception, but concluded that further studies are needed to understand how perceptivity depends on the underlying data. Earlier studies \cite{Lewis2012, Sedlmair2012} have also shown that the success of a quality metric largely depends on the underlying dataset. 

This paper suggests a set of QM that aims to guide visual exploration of structural missingness in large and high-dimensional datasets. To the best of our knowledge, this is the first attempt to utilise QM as an aid in exploring structural missingness to facilitate not only the management of missing values, but also to support the gaining of relevant insight from the missingness.

\section{Quality Metrics for Missingness Patterns} \label{sec:QM}
This section will describe QMs designed to support exploration and identification of missingness structures related to the three patterns defined by Johansson Fernstad \cite{Fernstad2019}. All of the suggested metrics are applicable to both numerical and categorical variables, either as is or with minor modifications to the algorithms.

The following notation will be used throughout the paper: a dataset $D$ includes $K$ variables and $N$ items; $\vec{d_i}$ is an item where $i = 1,2,...,N$; $\vec{d_j}$ and $\vec{d_k}$ are variables where $j,k = 1,2,...,K$; and $d_{i,j}$ is the value of item $i$ in variable $j$. For each variable, $D$ can be separated into two sets, where $D_{Mj}$ is the set of items for which values are missing in variable $j$ and $D_{Rj}$ is the set of items for which values are recorded in variable $j$.

\subsection{Amount Missing}
 Amount Missing (AM) refers to the relative amount of missing values, and supports understanding of the distribution of missing values across the dataset. Insight into AM structures can, for example, support identification of variables where the missingness may be extra difficult to deal with, or highlight subsets of data where conclusions drawn from the recorded values may be particularly unreliable due to the large amount of missing values. 

The QM for AM is focussed on the relative amount of missing values in a variable, and generates one value per variable in the range 0 to 1, with 1 corresponding to all values being missing. The metric, $Q_{AM}$, for variable $\vec{d_j}$ is defined in equation \ref{Eq:Q_AM}.

\begin{equation}\label{Eq:Q_AM}
Q_{AM}(\vec{d_j})=\frac{1}{N}\sum_{i=1}^{N}
\begin{cases}
    1 : i \in D_{Mj} \\
    0 : i \in D_{Rj}
  \end{cases}
\end{equation}

\subsection{Joint Missingness}
Joint Missingness (JM) is a multivariate or pairwise pattern that refers to the amount of data items that have missing values in more than one variable at the same time. The pattern may, e.g., occur in survey data where participants who do not to answer a specific question may also not answer other questions related to the first. Identification of JM can support discovery of issues in the data collection or pre-processing which cause missingness to propagate across the data, as well as recognition of data subsets where the missingness may need specific attention. Hence, the proposed JM metrics aim to measure the amount of data items that are concurrently missing in both variables of a variable pair. 

With a large number of missing values in each individual variable a certain amount of jointly missing values are expected to appear by chance, e.g., if two variables both have 50\% missing values individually it can be expected that 25\% will be jointly missing by chance. This expected JM can be defined as $E(\vec{d_j},\vec{d_k})=P(\vec{d_j}) \cdot P(\vec{d_k})$, where $P(\vec{d_j})$ and $P(\vec{d_k})$ are the probabilities that a value is missing in $\vec{d_j}$ and $\vec{d_k}$ respectively. When analysing missingness it is often of interest to identify structures that do not appear by chance, in particular to identify non-MCAR structures. Thus, while the magnitude of JM can be of relevance, the deviation from $E(\vec{d_j},\vec{d_k})$ may generally be of more interest. Based on this, three QM are suggested for identification of JM structures. The first metric, $Q_{JM_{mag}}$ (defined in equation \ref{Eq:Q_JMmag}), represents the relative magnitude of jointly missing values for pairs of variables, without taking the deviation from $E(\vec{d_j},\vec{d_k})$ into account.

\begin{equation}\label{Eq:Q_JMmag}
Q_{JM_{mag}}(\vec{d_j}, \vec{d_k})=\frac{1}{N}\sum_{i=1}^{N}
\begin{cases}
    1 & : i \in (D_{Mj} \cap D_{Mk})\\
    0 & : i \in (D_{Rj} \cup D_{Rk})
  \end{cases}
\end{equation}

The second metric, $Q_{JM_{dir}}$ (defined in equation \ref{Eq:Q_JMdir}) takes the direction of the deviation from expected JM into account. High (positive) values are assigned to variable pairs where the measured JM, defined as $P(\vec{d_j}, \vec{d_k})$, is higher than the expected JM (i.e., $P(\vec{d_j}, \vec{d_k}) > E(\vec{d_j},\vec{d_k})$), and low (negative) values to variable pairs where the measured JM is lower than the expected (i.e., $P(\vec{d_j}, \vec{d_k}) < E(\vec{d_j},\vec{d_k})$). 

\begin{equation}\label{Eq:Q_JMdir}
Q_{JM_{dir}}(\vec{d_j}, \vec{d_k})=P(\vec{d_j}, \vec{d_k}) - E(\vec{d_j},\vec{d_k})
\end{equation}

The third metric, $Q_{JM_{abs}}$ (defined in equation \ref{Q_JMabs}), focus only on the size of the deviation from the expected JM and does not take into account if the measured JM is higher or lower than expected. High values are hence assigned to variable pairs where the measured JM considerably deviates from the expected JM (i.e., $P(\vec{d_j}, \vec{d_k}) > E(\vec{d_j},\vec{d_k})$ or $P(\vec{d_j}, \vec{d_k}) < E(\vec{d_j},\vec{d_k})$), whereas values near 0 mean the measured JM is close to the expected JM and, hence, likely to occur by chance. 

\begin{equation}\label{Q_JMabs}
Q_{JM_{abs}}(\vec{d_j}, \vec{d_k})=|P(\vec{d_j}, \vec{d_k}) - E(\vec{d_j},\vec{d_k})|
\end{equation}

\subsection{Conditional Missingness}
Conditional Missingness (CM) is a pairwise pattern describing the relationship between items with missing values in one variable and their recorded value in other variables. It aims to describe patterns where the probability of missingness is conditional upon recorded values, and as such it supports understanding of relationships between missing and recorded. Investigation of CM can be useful for, e.g., recognising the cause of missingness, and can support decisions on how to best deal with the missingness, for example, if items with missing values in a variable $\vec{d_j}$ tend to have low recorded values in variable $\vec{d_k}$, then imputation of missing in $\vec{d_j}$ based on items with low values in $\vec{d_k}$ may be more reasonable than if the imputation was based on all items. 

The proposed metrics for CM consider pairwise relationships, similar to JM, and defines the missingness in one variable in relation to values in another variable. CM is directional and hence two metric values are associated with each variable pair. When a single value is required for a variable pair or a single variable, either the average, maximum or minimum value could be used. 

\begin{figure}[t]
\centering
\includegraphics[width=6cm]{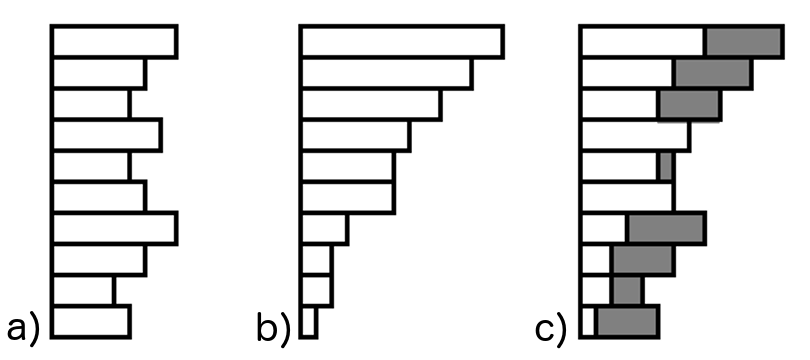}
\caption{Discrete probability distributions: a) distribution of all recorded values in $\vec{d_k}$, b) distribution in $\vec{d_k}$ of items that are missing in $\vec{d_j}$, c) the difference between the distributions in a) and b) highlighted in grey.}
\label{probDist}
\end{figure}

To examine if the missingness in variable $\vec{d_j}$ is conditional upon values in $\vec{d_k}$, the subset $D_{Mj}$ is considered in terms of its distribution in the recorded values of $\vec{d_k}$, and compared to the overall distribution of recorded values in $\vec{d_k}$. This enables investigation of whether the probability of an item being missing in $\vec{d_j}$ may be dependent on its value in $\vec{d_k}$, or if it is independent of it. The discrete probability distributions of missing and recorded values are suggested to examine distribution differences, using histograms of recorded values. The shape of the distribution in a histogram may be affected by the size of the histogram bins and, thus, an approach suggested by Shimazaki and Shinomoto \cite{Shimazaki2007} is used to identify the optimal number of bins, $b_j$, in each variable. For categorical variables each bin would instead correspond to a unique category. Two histograms of the probability distribution is created for each variable pair, as displayed in figure \ref{probDist}. The left histogram represents the probability distribution of all recorded values in $\vec{d_k}$ ($D_{Rk}$). The middle histogram represents the probability distribution in $D_{Rk}$ of the set of items that are missing in $\vec{d_j}$ ($D_{Rk} \cap D_{Mj}$). 

Two QM are suggested for CM, one based on differences in density distributions and the other based on entropy differences. Several other options could be considered, including QM based on the Kullback-Leibler Divergence \cite{Kullback1951}, Mutual Information or Conditional Entropy \cite{Shannon1948}. The first QM suggested here compares the difference in density distribution as displayed in figure \ref{probDist}c, where the difference between the two distributions is highlighted in grey. The metric is defined in equation \ref{Eq:Q_CMdid}, where $P_{D_{Rk}}(o)$ is the probability of a recorded value in $\vec{d_k}$ belonging to bin $o$, considering all recorded values in $\vec{d_k}$, and $P_{D_{Rk} \cap D_{Mj}}(o)$ is the probability that a recorded value in $\vec{d_k}$ belongs to bin $o$, considering only the set of items with missing values in $\vec{d_j}$. $b_k$ is the number of histogram bins in $\vec{d_k}$. The sum is finally divided by 2, the maximum possible difference, to obtain values in the range $[0, 1]$. High values are thus assigned when the distribution of missing values differ greatly from the overall distribution, which indicates that missing are not occurring at random and that there may exist a structural missingness relating to the values of variable $\vec{d_k}$. 

\begin{equation}\label{Eq:Q_CMdid}
Q_{CM_{DiD}}(\vec{d_j}, \vec{d_k}) = \frac{1}{2}\sum_{o=1}^{b_k}{| P_{D_{Rk}}(o) - P_{D_{Rk} \cap D_{Mj}}(o) |} 
\end{equation}

The entropy based metric, $Q_{CM_H}$, use the discrete probability distributions to compare the entropy of all recorded values in $\vec{d_k}$ with the entropy in $\vec{d_k}$ of the set of missing items in $\vec{d_j}$. The metric is defined in equation \ref{Eq:Q_CMh} 
, where $H_{D_{Rk}}$ and $H_{D_{Rk} \cap D_{Mj}}$ are the Shannon entropies \cite{Shannon1948} of the probability distributions in $\vec{d_k}$ of all recorded values in variable $\vec{d_k}$, and of items that are missing in variable $\vec{d_j}$. The entropy difference is divided by the maximum possible entropy to obtain a value range of $[0, 1]$, which for a discrete distribution is $log(b_k)$ where $b_k$ is the number of bins. Entropy provides a measure of the unpredictability in a variable and, hence, the entropy in variable $\vec{d_k}$ of missing values in $\vec{d_j}$ ($H_{D_{Rk} \cap D_{Mj}}$) can be seen as a measure of how predictable (or unpredictable) the missing values are based on the recorded values of $\vec{d_k}$. Similarly, the entropy in $\vec{d_k}$ of all items ($H_{D_{Rk}}$) can be seen as the overall predictability of $\vec{d_k}$. Difference in entropy, leading to a high $Q_{CM_H}$ value, can thus be interpreted as difference in predictability, indicating that missing values in $\vec{d_j}$ does not occur at random in relation to the overall entropy and in context of their values in $\vec{d_k}$.

\begin{equation}\label{Eq:Q_CMh}
Q_{CM_H}(\vec{d_j},\vec{d_k}) = \frac{| H_{D_{Rk}} - H_{D_{Rk} \cap D_{Mj}} |}{log(b_k)}
\end{equation}

\section{Visual Investigation of Missingness Structures} \label{sec:visualinvestigation}
\begin{figure}[tb]%
       \centering
       \subfloat[]
       [The relative amount missing in each variable is represented as a light blue block.]{
       	\includegraphics[width=3.6cm]{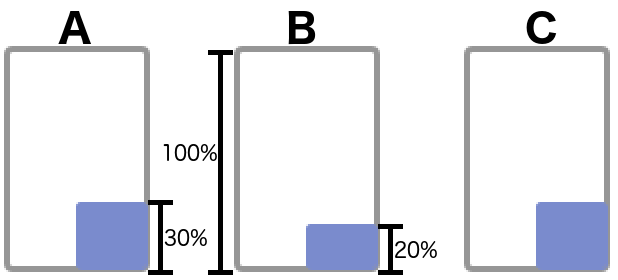}
       	\label{Fig:MissVisA}
       }%
       \qquad
       \subfloat[][Distribution of recorded items are represented as grey histograms, with $D$ being categorical.]{
       	\includegraphics[width=3.8cm]{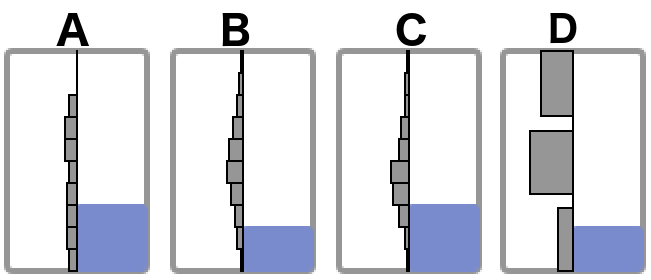}
       	\label{Fig:MissVisB}
       }%
       \qquad
       \subfloat[][The relative amount of jointly missing values with a selected variable is represented as a red block, with the selected variable ($C$) highlighted in red.]{
       	\includegraphics[width=3.7cm]{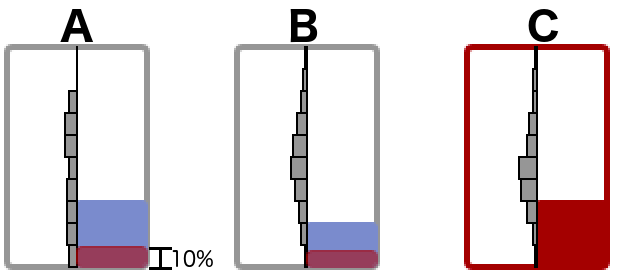}
       	\label{Fig:MissVisC}
       	}%
       \qquad
       \subfloat[][Red histograms represent the distribution of recorded values for items that have missing values in the selected variable ($C$).]{
       	\includegraphics[width=3.7cm]{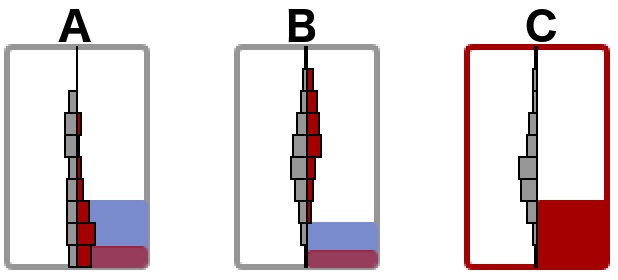}
       	\label{Fig:MissVisD}
       }
       \caption{The basic structure of MissiG for three or four variables \cite{fernstad2021explore}. Variable $C$ is selected in \ref{Fig:MissVisC} and \ref{Fig:MissVisD}, with patterns related to missing in $C$ represented by red in the other variables.}%
       \vspace{-2mm}
       \label{Fig:MissVisDescription}%
\end{figure}

This section will visually investigate synthetically generated missingness structures, described in section \ref{sec:dataGen}, to exemplify the use of the suggested QMs for variable layout, filtering, selection, and as the basis for visual representation of missingness structures (mainly through size and colour). Five visualization methods are used, as follows: 
\begin{itemize}
    \item \textbf{Barchart} representing $Q_{AM}$ (see figure \ref{fig:JM_teaser}).
    \item \textbf{Heatmap} with missing values represented by purple, and recorded values represented using a grey scale (see figure \ref{fig:JM_teaser}).
    \item \textbf{Parallel Coordinates} (PC) where missing values intersect below the axis, and polylines are coloured purple if the item has at least one missing value, and grey if all values are recorded (see figure \ref{fig:JM_teaser}).
    \item \textbf{MissiG} glyphs \cite{fernstad2021explore} that represent missingness structures of variables as described in figure \ref{Fig:MissVisDescription}.
    \item \textbf{Network} visualization using Cytoscape \cite{shannon2003cytoscape} where nodes represent individual variables, with node size corresponding to their $Q_{AM}$ value, and where the edge representations (width and colour) corresponds to QM values for variable pairs (see figure \ref{Fig:BCC_JM_NetworkJM}).
\end{itemize}

\noindent Larger and complementary versions of the figures in this section and the figures in section \ref{Sec:ICICLE} are available as supplemental material.

\subsection{Generation of Missingness Structures in Data}
\label{sec:dataGen}
To maintain control over the missingness structures in data, for the purpose of reviewing the utility of the proposed metrics, a publicly available dataset with no missing values ($BreastCancerCoimbra$ \cite{patricio2018using}) was used and modified to display different missingness structures. This dataset includes data from 106 patients (64 cancer patients and 52 healthy controls) for 10 variables, of which nine predictor variables and one binary dependent variable. 
Three new datasets (available as supplemental material), each representing varying levels of one of the three missingness structures (AM, JM, CM), were generated by replacing values in $BreastCancerCoimbra$ with $NaN$ strings, following the processes described below. 

\begin{table*}[tb]
    \caption{Pairwise JM patterns generated for the $BreastCancer_{JM}$ data.}
    \label{Table:BCC_JM}
    \scriptsize%
    \centering%
    \begin{tabu}{| l | c | c | c |}
\hline
\textbf{Variable pair (\% missing)} & \textbf{JM pattern} & $P(\vec{d_j}, \vec{d_k})$ &$E(\vec{d_j}, \vec{d_k})$\\
\hline
Age (32\%) ---	BMI (34\%) & $P(\vec{d_j}, \vec{d_k}) < E(\vec{d_j},\vec{d_k})$ & 3.6\% & 10.9\%\\
Glucose (33\%) ---	Insulin (33\%) & $P(\vec{d_j}, \vec{d_k}) < E(\vec{d_j},\vec{d_k})$ & 7.3\% & 10.9\%\\
HOMA (26\%) ---	Leptin (41\%) & $P(\vec{d_j}, \vec{d_k}) = E(\vec{d_j},\vec{d_k})$ & 10.7\% & 10.7\%\\
Adiponectin (46\%) ---	Resistin (33\%) & $P(\vec{d_j}, \vec{d_k}) > E(\vec{d_j},\vec{d_k})$ & 21.1\% & 15.2\%\\
MCP\_1 (46\%) --- Classification (50\%) & $P(\vec{d_j}, \vec{d_k}) > E(\vec{d_j},\vec{d_k})$ & 38.3.1\% & 23\%\\
\hline
\end{tabu}%
\end{table*}

\begin{table*}[tb]
\caption{Pairwise CM patterns generated for the $BreastCancer_{CM}$ data.}
\label{Table:BCC_CM}
\centering\begin{tabu}{| l | c | c | c |}
\hline
\textbf{Variable pair (\% missing)} & \textbf{Condition range type} & \textbf{Condition range} & \textbf{Strength level} \\
\hline
Age (28\%) ---	BMI (0\%) & \textit{Medium values} & 24.24 - 29.77  & \textit{Low-CM} \\
Glucose (15\%) ---	Insulin (0\%) & \textit{High values} & 8.56 - 58.46 & \textit{Medium-CM} \\
HOMA (26\%) ---	Leptin (0\%) & \textit{Medium values} & 14.74 - 29.22 & \textit{High-CM}  \\
Adiponectin (25\%) ---	Resistin (0\%) & \textit{Medium values} & 8.04 - 15.68 & \textit{Low-CM} \\
MCP\_1 (14\%) --- Classification (0\%) & \textit{Low values} & 1.0 - 1.0 & \textit{Medium-CM} \\
\hline
\end{tabu}
\end{table*}

\noindent \textbf{AM:} For the AM version of the dataset ($BreastCancer_{AM}$), a missingness level between 0\% and 50\% where randomly chosen for each variable. Corresponding number of values were then randomly replaced by a $NaN$ string for each variable, resulting in a dataset with a total of 273 missing values (30.8\%), with between 7.8\% and 46.6\% missing in each variable.

\noindent \textbf{JM:} Three different types of JM patterns were implemented for $BreastCancer_{JM}$: 
\begin{enumerate}[noitemsep]
    \item Actual JM corresponds to the expected: $P(\vec{d_j}, \vec{d_k}) = E(\vec{d_j},\vec{d_k})$.
    \item Actual JM is higher than the expected: $P(\vec{d_j}, \vec{d_k}) > E(\vec{d_j},\vec{d_k})$.
    \item Actual JM is lower than the expected: $P(\vec{d_j}, \vec{d_k}) < E(\vec{d_j},\vec{d_k})$.
\end{enumerate}
\noindent To achieve clear patterns a relatively high level of missingness was randomly assigned to each variable, ranging from 20\% to 50\% missing. The 10 variables of the dataset where then treated as 5 individual variable pairs, as displayed in table \ref{Table:BCC_JM}, with $E(\vec{d_j},\vec{d_k})$ calculated based on the variables individual missingness levels. The pairs were randomly assigned to one of the three JM patterns and an actual JM value, $P(\vec{d_j},\vec{d_k})$, was assigned based on the pattern. The corresponding number of values were then replaced by a $NaN$ strings such that they fulfil both the individual and pairwise missingness for the variables.

\noindent \textbf{CM:} For $BreastCancer_{CM}$ the 10 variables of the dataset were treated as 5 individual variable pairs, where one of the variables in the pair includes missing values ($\vec{d_j}$) which are conditional upon values in the other variable ($\vec{d_k}$), hence, missing values were only generated for $\vec{d_j}$. 
The amount missing for $\vec{d_j}$ was randomly set to a value between 10\% and 33\% for each variable pair, as displayed in table \ref{Table:BCC_CM}, and a condition range was randomly assigned such that the missingness was conditional upon: 

\begin{enumerate}[noitemsep]
\item \textit{Low values}: the lowest third of recorded values in $\vec{d_k}$.
\item \textit{Medium values}: recorded values within the second third in $\vec{d_k}$.
\item \textit{High values}: the top third of recorded values in $\vec{d_k}$
\end{enumerate}

\noindent Finally, three strength levels of CM were used, defined as: 

\begin{enumerate}[noitemsep]
\item \textit{Low-CM}: 30\% of missing values are within the conditional range.
\item \textit{Medium-CM}: 60\% of missing values are within the conditional range.
\item \textit{High-CM}: 90\% of missing values are within the conditional range.
\end{enumerate}

\noindent Recorded values in $\vec{d_j}$ were then replaced by $NaN$ strings to fulfil the set conditions for each variable pair.

Sections \ref{sec:investig_AM} -- \ref{sec:investig_CM} will investigate the three missingness structures using the above described datasets. This is then followed by a case study in section \ref{Sec:ICICLE} were the proposed metrics are utilised to explore missingness structures in a high-dimensional health dataset that contains a large amount of missing values.

\subsection{Visual Investigation of Amount Missing}
\label{sec:investig_AM}
\begin{figure}[t]%
        \centering
       \subfloat[][Randomly ordered variables with $BMI$ and $Leptin$ underlined.]{
       	\includegraphics[width=8.5cm]{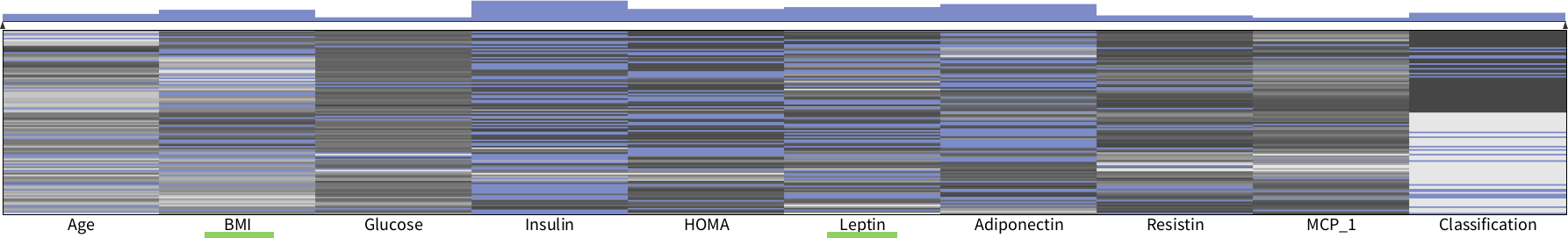}
       	\label{Fig:BCC_AM_noOrder_HM}
       }%
       \qquad
       \subfloat[][Visualization of the full $BreastCancer_{AM}$ dataset, with variables ordered from left to right by $Q_{AM}$ and $Leptin$ and $BMI$ underlined.]{
       	\includegraphics[width=8.5cm]{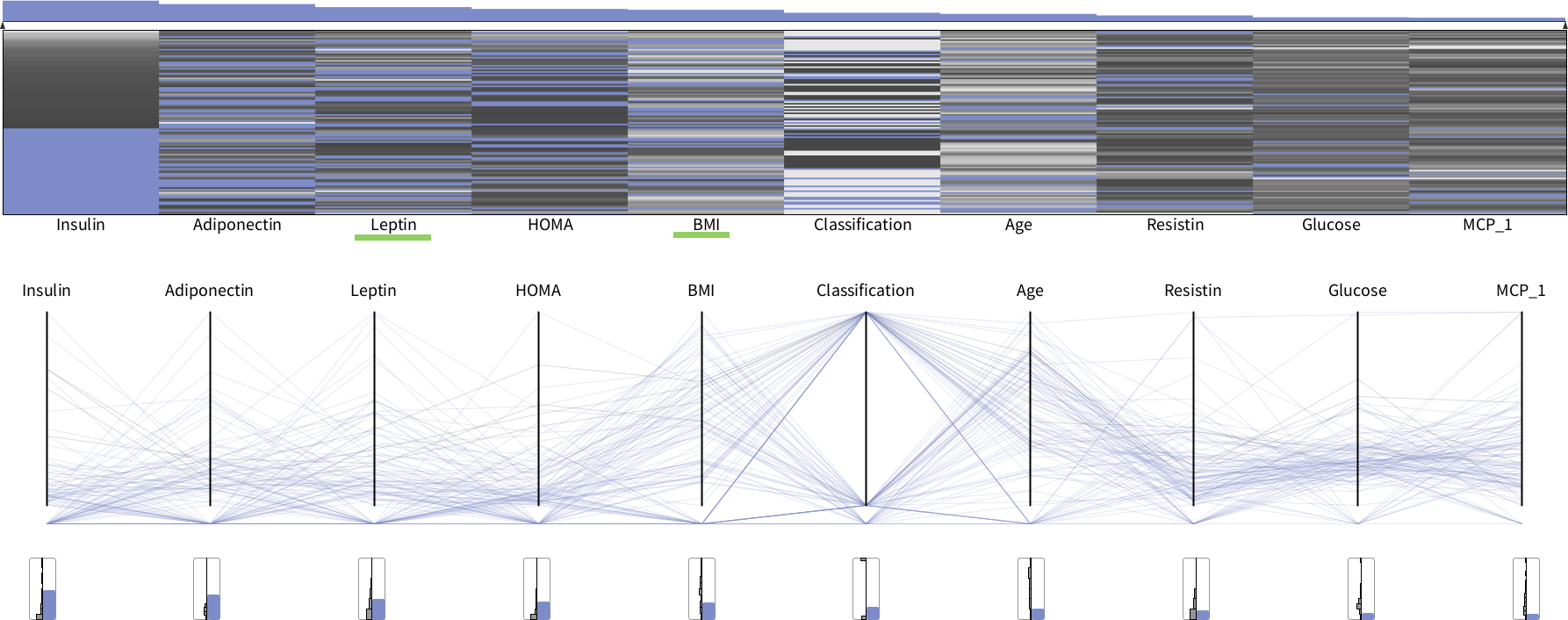}
       	\label{Fig:BCC_AM_unfiltered}
       }%
       \qquad
       \subfloat[][Visualization of $BreastCancer_{AM}$, filtered to display the five variables with highest $Q_{AM}$ in PC and MissiG. Items with missing values in the leftmost variable ($Insulin$) are highlighted in red.]{
       	\includegraphics[width=8.5cm]{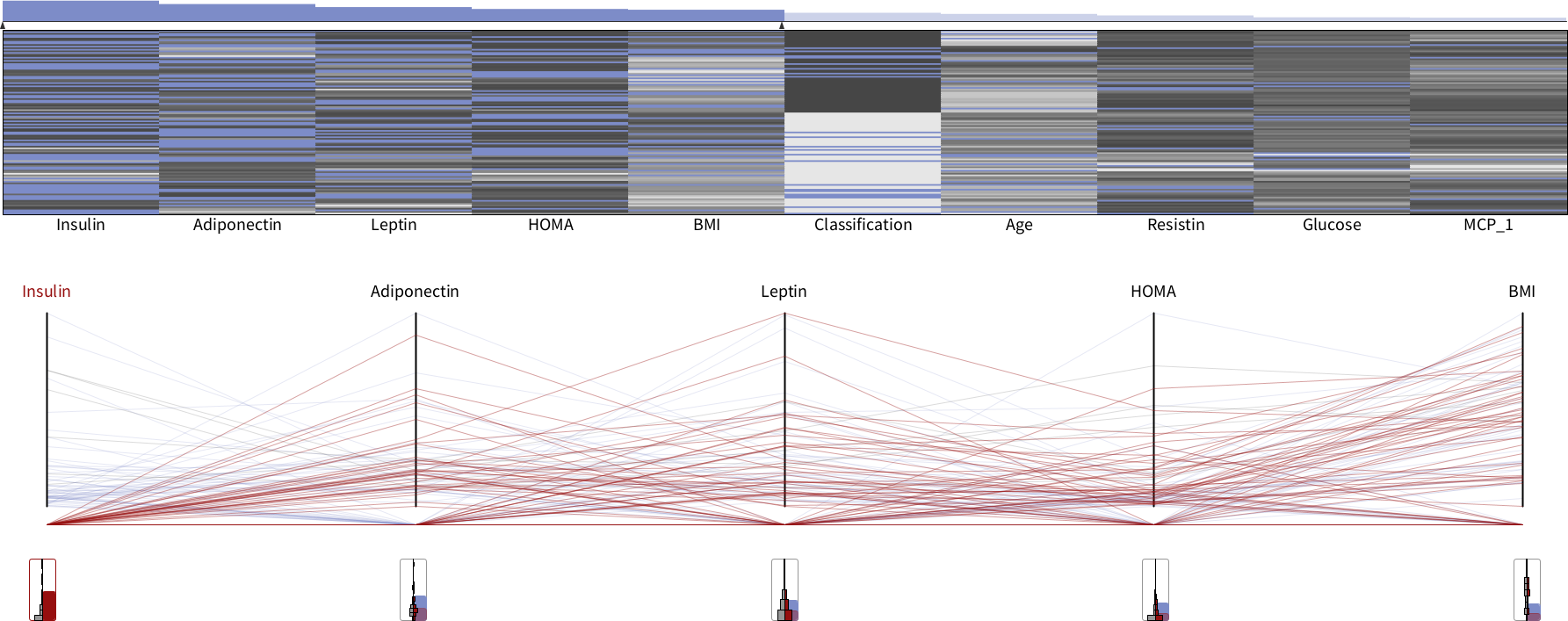}
       	\label{Fig:BCC_AM_filtered}
       	}
       \caption{Visualization of $BreastCancer_{AM}$. Missing values are represented below the axes in PC, and MissiG glyphs are used to display further missingness structures.}%
       \vspace{-2mm}
       \label{Fig:BCC_AM_2}%
\end{figure}

The calculated $Q_{AM}$ values for the variables in the $BreastCancer_{AM}$ dataset range from 0.078 to 0.466.
Figure \ref{Fig:BCC_AM_noOrder_HM} displays the dataset visualized using a combination of heatmap and barchart, where purple colour and height of bars represent missingness structures and variables (i.e. columns) are randomly ordered. 
Figure \ref{Fig:BCC_AM_unfiltered} displays the same data with variables ordered by $Q_{AM}$ from left to right and heatmap rows ordered by $Insulin$ (leftmost column), including a PC and MissiG representation of the same data at the bottom. It is clear from the heatmaps and barcharts that there is variation in amount missing between variables, indicating that values are not missing completely at random. The QM based ordering further facilitate identification and comparison of $Q_{AM}$ structures as, for example, the difference between $Leptin$ and $BMI$ is easier to perceive in the barchart in \ref{Fig:BCC_AM_unfiltered} (third and fifth variable, underlined in green) than in the unordered figure \ref{Fig:BCC_AM_noOrder_HM} (sixth and second variable, underlined in green).

In figure \ref{Fig:BCC_AM_filtered}, $Q_{AM}$ has been used for filtering to display only the five variables with highest amount missing in PC and MissiG. The highlighting in red of items with missing values in $Insulin$ along with the row ordering in the heatmap does not indicate any obvious JM or CM structures, since in the heatmap the colour distribution in the block of rows with missing values in $Insulin$ are similar to the section with recorded values, and since the red lines in the PC are equally distributed to non-red lines.

\subsection{Visual Investigation of Joint Missingness}
\label{sec:investig_JM}
\begin{figure}[!t]
\centering
\includegraphics[width=8.5cm]{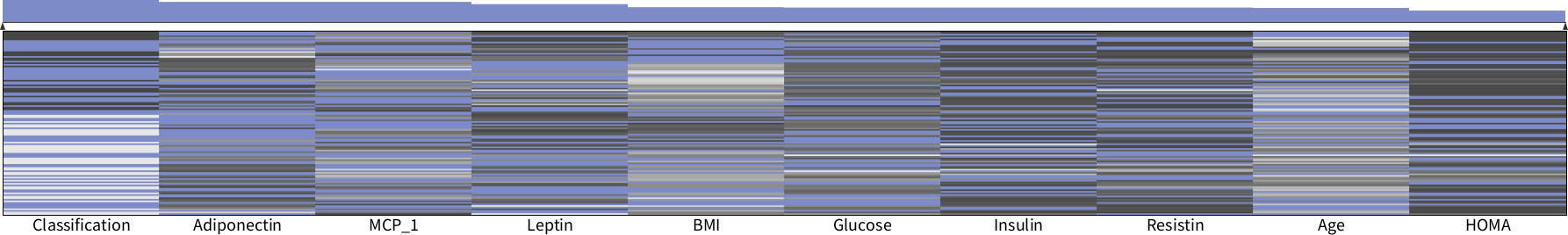}
\caption{$BreastCancer_{JM}$ displayed in heatmap and barchart ordered from left to right by $Q_{AM}$.}
\label{Fig:BCC_JM_AMorder}
\vspace{-4mm}
\end{figure}

$Q_{AM}$ values for the variables in the $BreastCancer_{JM}$ dataset 
ranges from 0.259 to 0.5. Table \ref{Table:BCC_JM_JMCM} provides statistical summaries for the three JM metrics for the dataset. 
Visualization of the dataset using barchart and heatmap (figure \ref{Fig:BCC_JM_AMorder}) confirms that all variables contain a comparably high number of missing values. 

\begin{table}[!t]
\caption{Statistical summaries of Joint Missingness metrics for variable pairs in $BreastCancer_{JM}$}
\label{Table:BCC_JM_JMCM}
\centering\begin{tabu}{| l | c  c  c  c  c|}
\hline
& \textit{Min}	& \textit{Med} 	& \textit{Avg} & \textit{Max} & \textit{Std} \\ 
\hline
$Q_{JMmag}$ & 0.034 & 0.138	& 0.140	& 0.379	& 0.056 \\
$Q_{JMabs}$ & 0	& 0.012 &	0.021	& 0.151	& 0.026 \\
$Q_{JMdir}$ & -0.073	& 0.002	& 0.003	& 0.151	& 0.033 \\
\hline
\end{tabu}
\vspace{-3mm}
\end{table}

\begin{figure}[t]%
       \centering
       \subfloat[][Layout weighted by $Q_{JMabs}$, edge width corresponding to $Q_{JMabs}$, and colour by $Q_{JMdir}$, with green for negative values and purple for positive.]{
       \includegraphics[width=7.5cm]{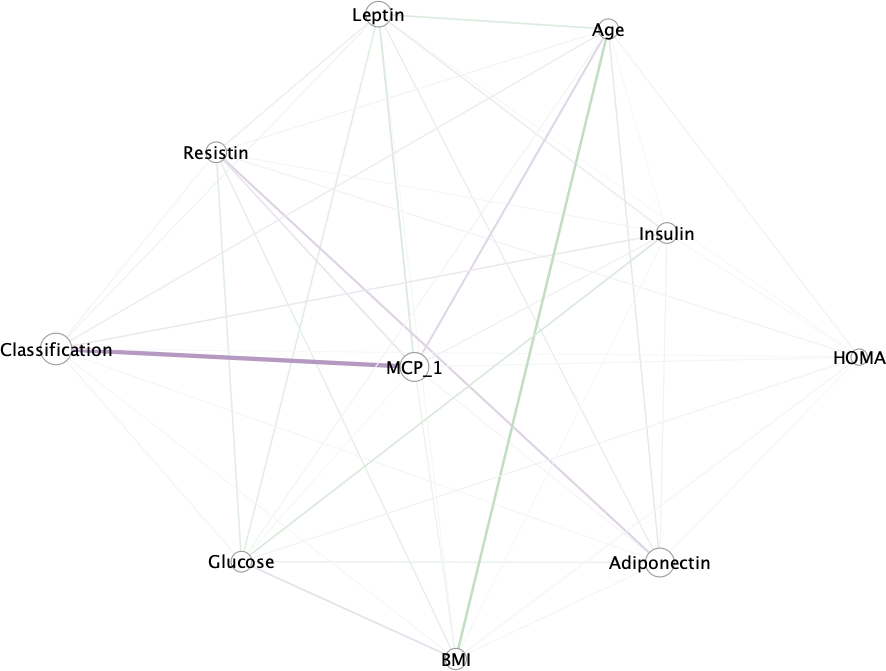}
       	\label{Fig:BCC_JM_Network_abs}
       }%
       \qquad
       \subfloat[][Layout weight, edge width and colour by $Q_{JMmag}$, with light blue for low values and dark green for high.]{
       	\includegraphics[width=7.5cm]{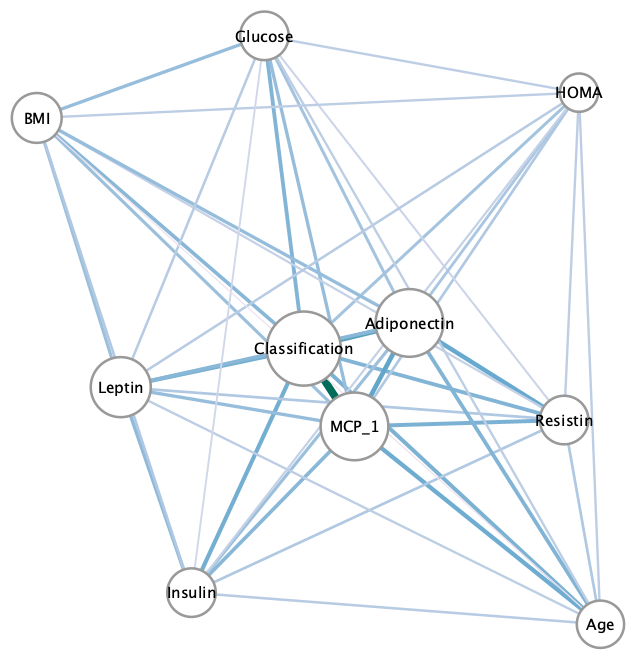}
       	\label{Fig:BCC_JM_Network_mag}
       	}
       \caption{Network visualization of $BreastCancer_{JM}$ with layout and visual appearance based on JM metrics.
       }%
       \vspace{-5mm}
       \label{Fig:BCC_JM_NetworkJM}%
\end{figure}

\begin{figure}[t]%
       \centering
       \subfloat[]
       [$MCP\_1$ (second from right) is selected. Arc width indicate higher $Q_{JMmag}$ with $Classification$ (rightmost) than with other variables, and height of red MissiG bar for $Classification$ indicate higher than expected JM.]{
       \includegraphics[width=8.5cm]{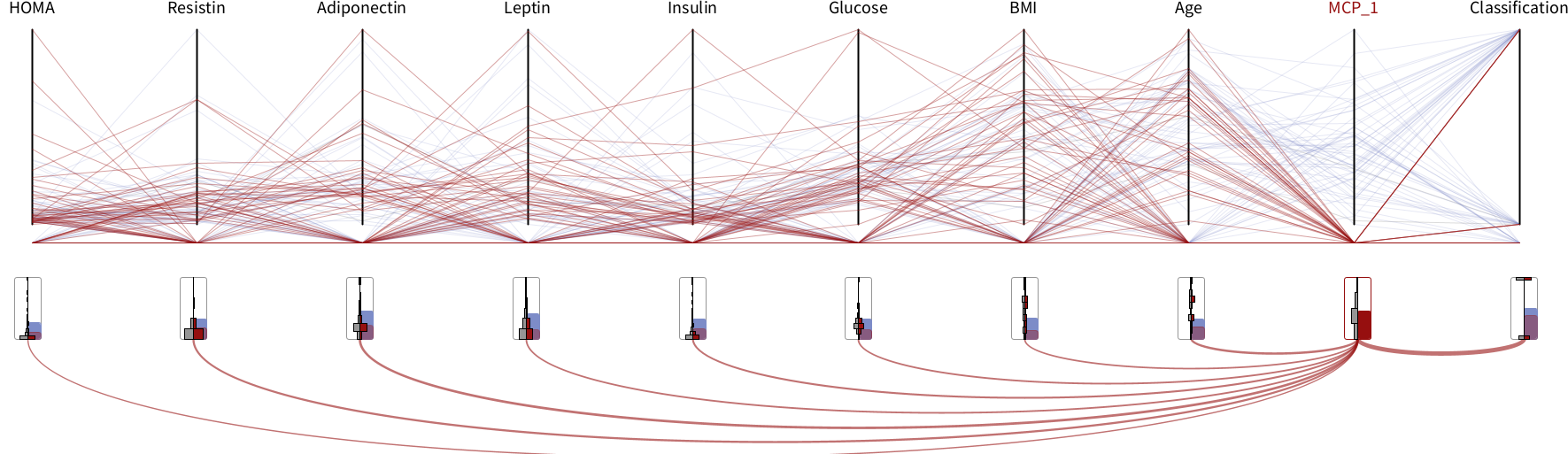}
       	\label{Fig:BCC_JM_PC_high}
       }%
       \qquad
       \subfloat[][$BMI$ (fourth from right) is selected. Arc width indicate lower $Q_{JMmag}$ with $Age$ (third from left) than with other variables, and low height of red MissiG bar for $Age$ indicate lower than expected JM.]{
       	\includegraphics[width=8.5cm]{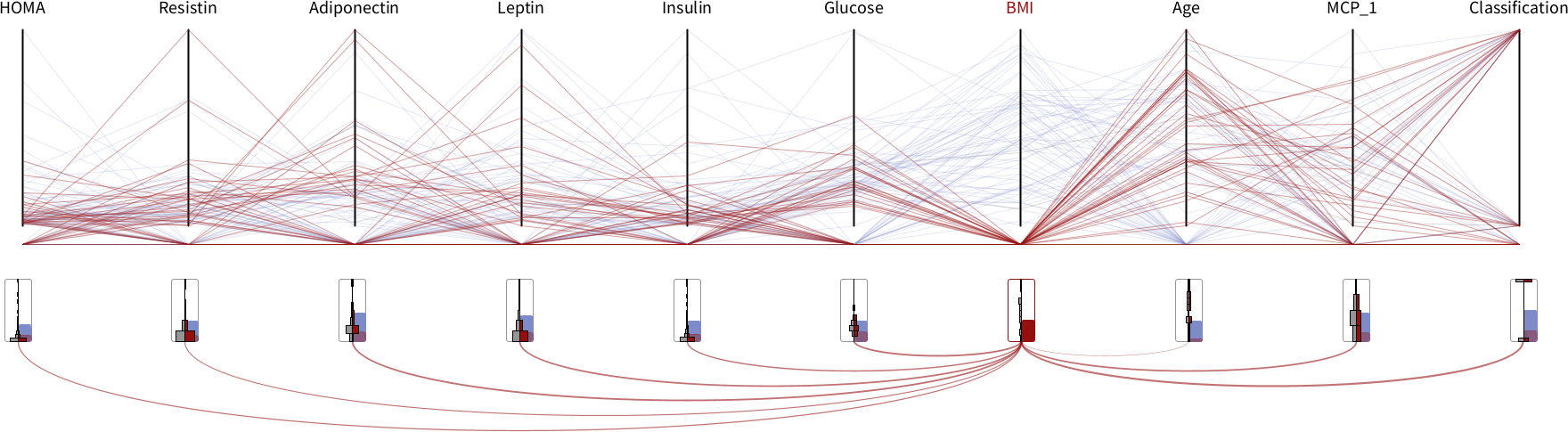}
       	\label{Fig:BCC_JM_PC_low}
       	} %
       \caption{$BreastCancer_{JM}$ visualized using PC and MissiG, with variables ordered by pairwise $Q_{JMabs}$ and width of arcs corresponding to pairwise $Q_{JMmag}$ with the selected variable.
       }%
       \vspace{-2mm}
       \label{Fig:BCC_JM_PC}%
\end{figure}

The datasets JM structures are initially investigated through network visualizations (figure \ref{Fig:BCC_JM_NetworkJM}) where node sizes corresponds to the variables $Q_{AM}$. In figure \ref{Fig:BCC_JM_Network_abs} the network use a spring embedded layout with $Q_{JMabs}$ as edge weight. Deviations from expected JM are further emphasized through edge width based on $Q_{JMabs}$, and edge colouring by $Q_{JMdir}$ with green representing negative values (i.e. $P(\vec{d_j}, \vec{d_k}) < E(\vec{d_j},\vec{d_k})$) and purple representing positive values (i.e. $P(\vec{d_j}, \vec{d_k}) > E(\vec{d_j},\vec{d_k})$). Through this, unexpectedly high or low JM of variables can be easily identified. For example, figure \ref{Fig:BCC_JM_Network_abs} reveals a strong, higher than expected JM between $Classification$ and $MCP\_1$, visible from the thicker purple edge. Similarly, the relatively thick green edge between $BMI$ (bottom node) and $Age$ (top-right node) indicates a lower than expected JM. 

These structures can be explored further using PC and MissiG, as in figure \ref{Fig:BCC_JM_PC} where the variables are ordered by pairwise $Q_{JMabs}$ using the ordering algorithm suggested by Artero et al.\ \cite{artero2006enhanced} and also described in \cite{Johansson2009}. The variable pair with the highest $Q_{JMabs}$ forms the basis of the ordering and the remaining variables are ordered one by one on the left or right side based on their $Q_{JMabs}$ with the outermost ordered variables, leading to variables with high $Q_{JMabs}$ located in close proximity of each other. $MCP\_1$ is selected in figure \ref{Fig:BCC_JM_PC_high}, with structures and items with missing values in $MCP\_1$ highlighted in red. The width of the red arcs represent the number of jointly missing values (i.e. $Q_{JMmag}$) and, hence, the comparably thick red arc between $MCP\_1$ and $Classification$ indicate a higher $Q_{JMmag}$ between these two variables than between $MCP\_1$ and other variables, which agrees with findings in figure \ref{Fig:BCC_JM_NetworkJM}. 

The lower than expected JM between $BMI$ and $Age$ is investigated further in figure \ref{Fig:BCC_JM_PC_low} (fourth and thrid axis from right), where $BMI$ and items and structures related to missing values in $BMI$ are highlighted in red. As expected, the arc between the two variables is thin, indicating a low $Q_{JMmag}$. Looking at the height of the red and blue bars in the MissiG representation, around one third of values are missing in both $BMI$ and $Age$. If missingness in the variable pair was random, it would be expected that around 10\% (one third of one third) was jointly missing. However, the height of the red JM bar in the $Age$ glyph (third from right) appear to correspond to considerably less than one third of the blue bar, hence visually confirming the lower than expected JM.

Figure \ref{Fig:BCC_JM_Network_mag} displays another network layout of the same dataset with edge weight, width and colour based on $Q_{JMmag}$. Light blue edges correspond to variable pairs with a low magnitude of jointly missing values, and green edges correspond to pairs with high joint missingness. As expected, $Classification$ and $MCP\_1$ are in close proximity of each other and have a thick green edge, as a result of their high $Q_{JMmag}$ value. The network furthermore reveal relatively high $Q_{JMmag}$ between $Classification$ and, e.g., $Leptin$ and $Adiponectin$. This may be an effect of the individual high amount missing in these variables resulting in high $Q_{JMmag}$, as the values does not appear to differ much from the expected JM according to edge width and colour in figure \ref{Fig:BCC_JM_Network_abs}.

\subsection{Visual Investigation of Conditional Missingness}
\label{sec:investig_CM}
$Q_{AM}$ values for the variables in the $BreastCancer_{CM}$ dataset 
ranges from 0 to 0.284. As a result of the pairwise approach when generating the CM structures (described in section \ref{sec:dataGen}) half of the variables have no missing values.
Table \ref{Table:BCC_CM_CM} provides statistical summaries for the two CM metrics, based on all variable pairs in the dataset.

\begin{table}[!t]
\caption{Statistic summaries of Conditional Missingness metrics for variable pairs in $BreastCancer_{CM}$}
\label{Table:BCC_CM_CM}
\centering\begin{tabu}{| l | c  c  c  c  c |}
\hline
& \textit{Min}	& \textit{Med} 	& \textit{Avg} & \textit{Max} & \textit{Std} \\ 
\hline
$Q_{CMdid}$	& 0 & 0.140 & 0.148 & 0.350 &	0.116 \\
$Q_{CMh}$ & 0 &	0.058 &	0.074 & 0.275 &	0.078 \\
\hline
\end{tabu}
\vspace{-3mm}
\end{table}

\begin{figure}[!t]
\centering
\includegraphics[width=6.5cm]{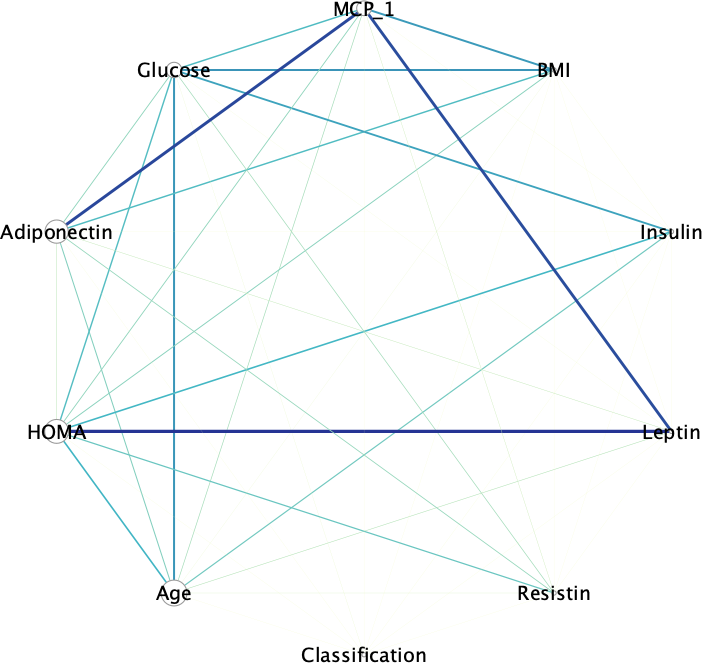}
\caption{Network visualization of $BreastCancer_{CM}$ using a circular layout. Edge width corresponds to $Q_{CMh}$, and colour by $Q_{CMh}$ with yellow representing low values and blue representing high values.}
\label{Fig:BCC_CM_CMnetwork}
\end{figure}

\begin{figure}[!t]
\centering
\includegraphics[width=8.5cm]{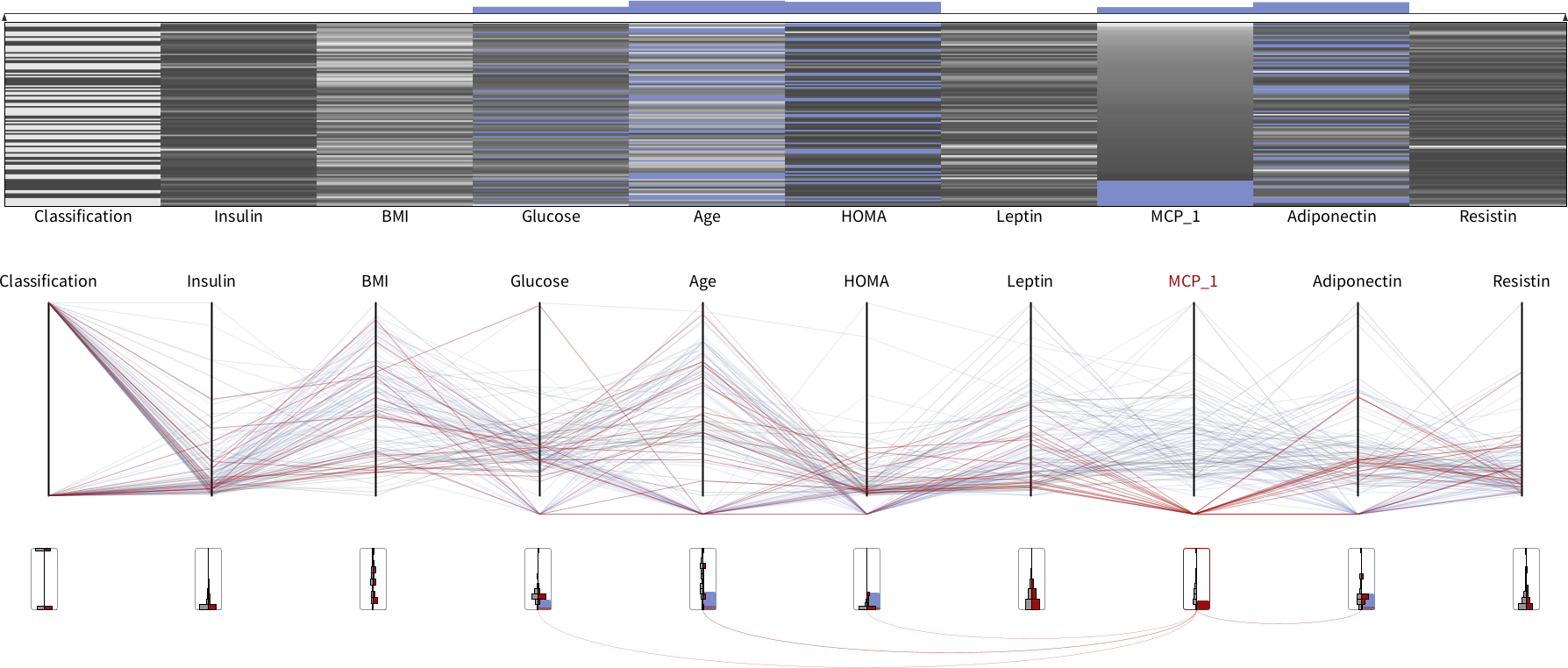}
\caption{Visualization of $BreastCancer_{CM}$ ordered by $Q_{CMh}$. $MCP\_1$ is selected in all views, and items with missing values in $MCP\_1$ are highlighted in red in PC and MissiG.}
\label{Fig:BCC_CM_CMorder}
\end{figure}

A network visualization (figure \ref{Fig:BCC_CM_CMnetwork}) is used for initial exploration of CM patterns in $BreastCancer_{CM}$. 
A circular layout is used with ordering based on $Q_{AM}$ and edge width and colour based on $Q_{CMh}$ values of the variable pairs, with yellow representing low values and blue representing high values. The figure reveals strong CM patterns between $MCP\_1$ (top node) and $Adiponectin$ (left) as well as with $Leptin$ (lower right), and between $HOMA$ (lower left) and $Leptin$. 

The structures involving $MCP\_1$ is further investigated using heatmap, PC and MissiG (figure \ref{Fig:BCC_CM_CMorder}), where variable pairs are ordered by $Q_{CMh}$ using the approach by Artero et al. \cite{artero2006enhanced}. $MCP\_1$ is selected in all views and items with missing values in $MCP\_1$ are highlighted in red across variables in PC and MissiG. The distribution of red lines in the PC indicate a relationship between missing items in $MCP\_1$ and low values recorded in $Leptin$, as well as low values in $Adiponectin$.

\subsection{Summary}
This section has demonstrated how the suggested QM can be used to identify missingness structures in datasets with structural missingness. It has also expemplified a variety of ways in which QM can be used as part of visual investigation to extract and emphasize patterns of interest, including variable ordering, subset selection, network layout, and visual representation. Furthermore, algorithms for generation of synthetically structured missingness has been suggested, to facilitate controlled testing of methods for missing data analysis. 

\section{ICICLE Case Study}
\label{Sec:ICICLE}
This case study provides an exemplar of how the proposed QM can support visual exploration to gain insight into structural missingness in a real-life walking monitoring dataset from the Incidence of Cognitive Impairment in Cohorts with Longitudinal Evaluation (ICICLE) \cite{lord_cognition_2014, yarnall_characterizing_2014}. 
Mobility monitoring for health purposes can greatly improve the understanding of certain pathologies. Sensor technologies enable recording of mobility metrics but analysis can often face data quality challenges such as dealing with large amounts of missing data. The causes of missingness can range from technical issues (e.g., sensor faults), to patient or health related issues (e.g., inability to participate due to declining health). Missingness caused by health issues is particularly important to understand for longitudinal mobility monitoring studies. 

Due to the nature of statistical analysis, where missing values are commonly removed, patients with incomplete records will often have less impact on analysis and results will be driven by the less impaired patients whose data are recorded across the whole study. Nonetheless, patients with missing values caused by declining health may often be the most important to focus clinical effort on, as they experience more problems and are worse impaired.

The visual investigation presented in this case study was carried out in collaboration with experts in bio-engineering and digital health, who are also co-authors of this paper. The sole purpose of the analysis is to demonstrate the potential of the suggested methods for identifying and exploring structural missingness. There is no intention to highlight or reveal any medical findings, and all such reflections are purely speculative.

\subsection{The data}
The ICICLE dataset contains data on patients with Parkinson's disease and healthy controls. It includes quantified measures of gait characteristics and walking behaviour (e.g., bout time, number of steps, mean step length, step length asymmetry, etc.), collected longitudinally in free living conditions and lab environment over six years on an eighteen-month interval basis (T0, T18, T36, T54, T72) using sensors on the lower back. It also include contextual patient data (e.g., disease group, sex, height, weight, etc.) and cognitive efficiency scores. In total, the dataset contains data for 206 attributes and 303 patients, and has a total of 56\% (34,947) missing values.

\subsection{Amount Missing}
\begin{figure}[t]
\centering
\includegraphics[width=8.5cm]{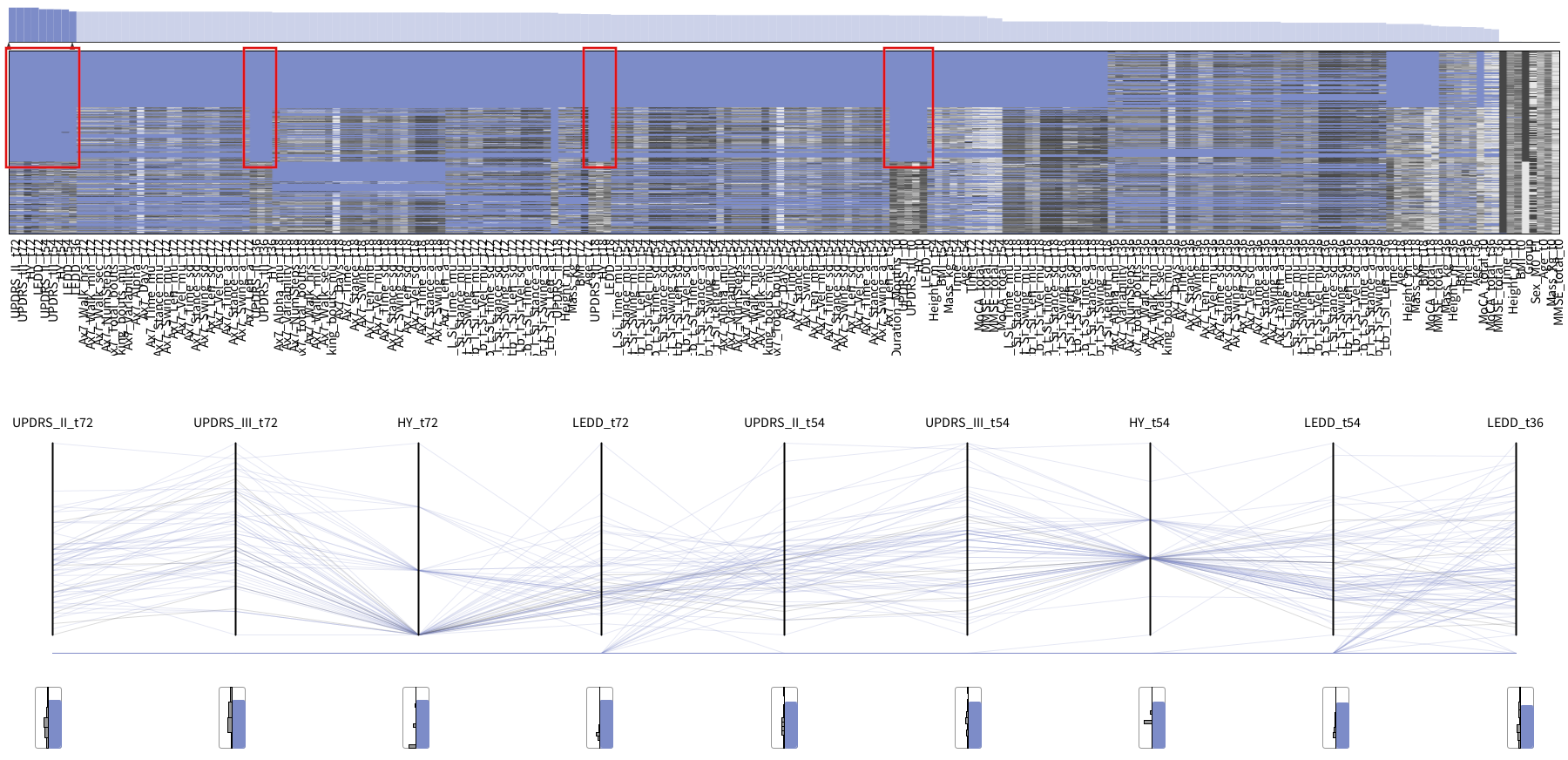}
\caption{The 206 attributes of the ICICLE dataset ordered by $Q_{AM}$ in the barchart and heatmap, with highest $Q_{AM}$ to the left. Selection of the 9 attributes with highest $Q_{AM}$ is reflected in PC and MissiG at the bottom.}
\vspace{-3mm}
\label{Fig:ICICLE_AMorder_HMPC}
\end{figure}

Initial visual inspection of the full dataset (figure \ref{Fig:ICICLE_AMorder_HMPC}) with attributes ordered from left to right by $Q_{AM}$ confirms, through the amount of purple cells in the heatmap, that a large part of the values are missing. The heatmap furthermore reveals several purple blocks of missing values across groups of attributes and patients. 

Thresholding by $Q_{AM}$ in the barchart at the top selects a block of attributes that holds the highest amount of missing values, which are displayed in the PC and MissiG at the bottom of the figure. This $Q_{AM}$ guided attribute selection reveals a set of attributes of the types $UPDRS$\textunderscore$II$, $UPDRS$\textunderscore$III$, $HY$, and $LEDD$, collected at later time points (T36, T54, T72). These are all clinical measures related to Parkinson's disease, and further investigation reveals that the records in the large purple block for these attributes in the heatmap (enclosed by red border) are all from control participants for which these measures were not recorded. Additionally, three similar blocks of missing values are visible across the heatmap, enclosed by red borders in figure \ref{Fig:ICICLE_AMorder_HMPC}. The attributes of these blocks correspond to similar clinical measures at different time points, with later time points located further to the left, confirming that the total amount missing for these attributes increases with time alongside the block of control records. This finding may be explainable by missingness due to declining health, where clinical mobility measures cannot be recorded due to some patients' inability to participate in the later part of the study.

\subsection{Joint Missingness}
With a high amount of missing values across a majority of attributes, it is expected that the magnitude of jointly missing values, $Q_{JMmag}$, will be high for most variable pairs. The deviation from expected joint missing may, hence, provide more valuable insights. 

\begin{figure*}[h]
  \centering
  \includegraphics[width=0.9\linewidth]{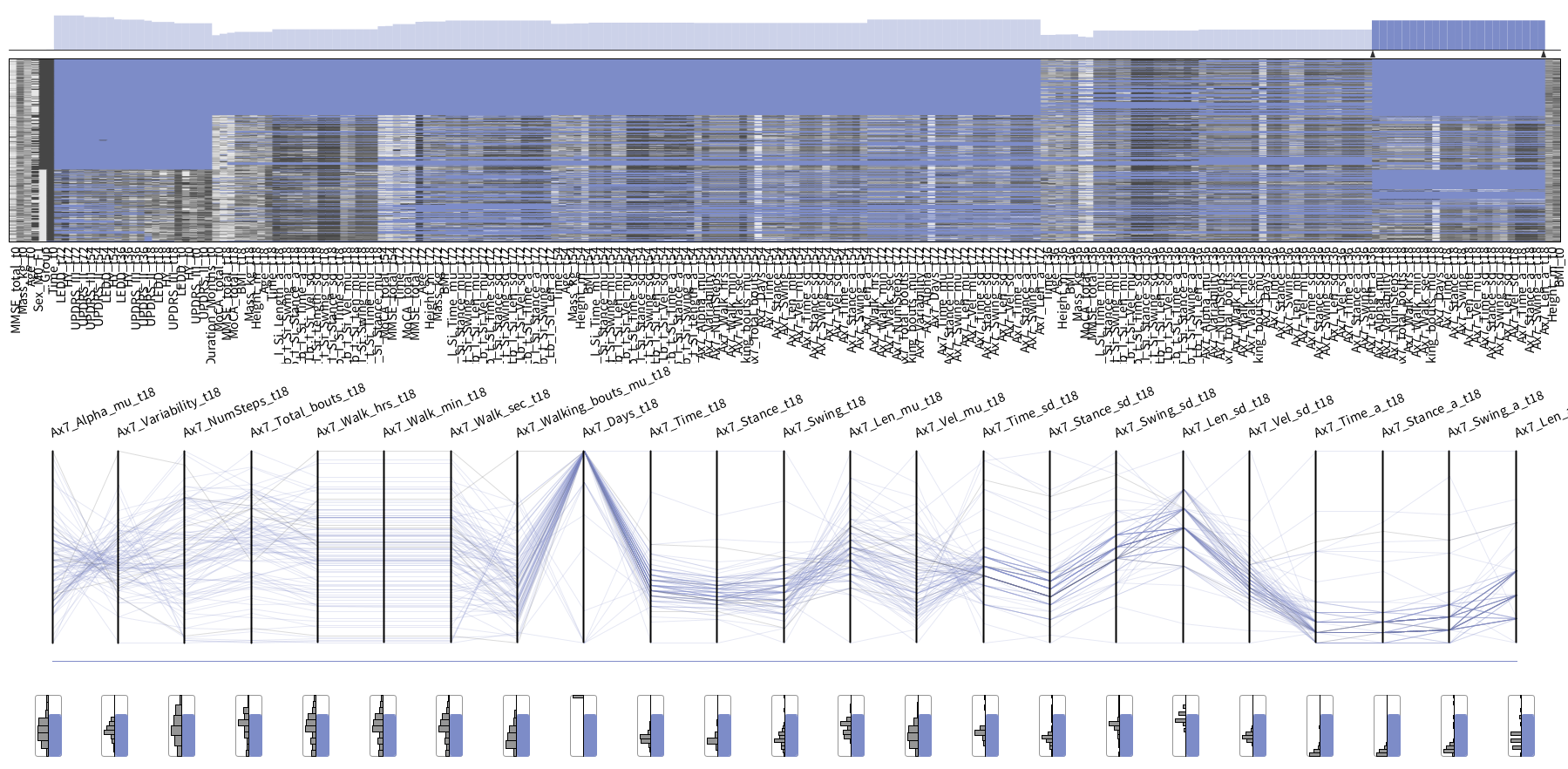}
  \caption{%
  Visual analysis of Joint Missingness structures in the ICICLE dataset. The 206 attributes are ordered by the $Q_{JMabs}$ Quality Metric in the barchart and heatmap at the top. 23 attributes with high pairwise $Q_{JMabs}$ are selected in the barchart and displayed in the Parallel Coordinates and MissiG visualization at the bottom.
  }
  \label{fig:JM_teaser}
\end{figure*}

Figure \ref{fig:JM_teaser} displays the full dataset with variables ordered by $Q_{JMabs}$ in the heatmap and barchart, using the approach by Artero et al.\ \cite{artero2006enhanced}. This ordering enables identification of groups of attributes with unexpectedly high or low joint missingness, i.e., high $Q_{JMabs}$. A group of attributes with very high amount of missing and jointly missing values are instantly detected in the left part of the heatmap. It is notable that this large block of purple cells can be associated with the black cells in the 5th column ($Group$), which corresponds to healthy control participants. The attributes of the block are largely the same that were revealed in figure \ref{Fig:ICICLE_AMorder_HMPC} (mainly of the type $UPDRS$\textunderscore$II$, $UPDRS$\textunderscore$III$, $HY$, and $LEDD$) across all time points of the study. Hence, this unexpectedly high JM can be explained by the data collection process where some metrics were not recorded for healthy participants.  

A separate pattern on the right hand side of the heatmap in Fig \ref{fig:JM_teaser} indicate a partly different structure through a group of attributes with high joint missingness (revealed through blocks of purple cells across rows and attributes), where a comparably large part of the jointly missing values occur also for participants with Parkinson's disease (white cells in column 5 from the left). A subset selection of these attributes, through thresholding in the barchart, is reflected in the PC and MissiG visualization at the bottom of the figure. This $Q_{JMabs}$ guided selection reveals a group of attributes which all represent measures of the $Ax7$ type at the 18 month time point. It is confirmed that also this missingness structure can be linked to data collection, since $Ax7$ are 7-day measures of free living activity which were recorded only for a subset of participants and, hence, values are missing for those not taking part. Furthermore, a different sensor was used at some time points and the change of device meant that activity measures for certain groups of participants were not measured equally across all time points. 

\begin{figure}[t]
\centering
\includegraphics[width=8.5cm]{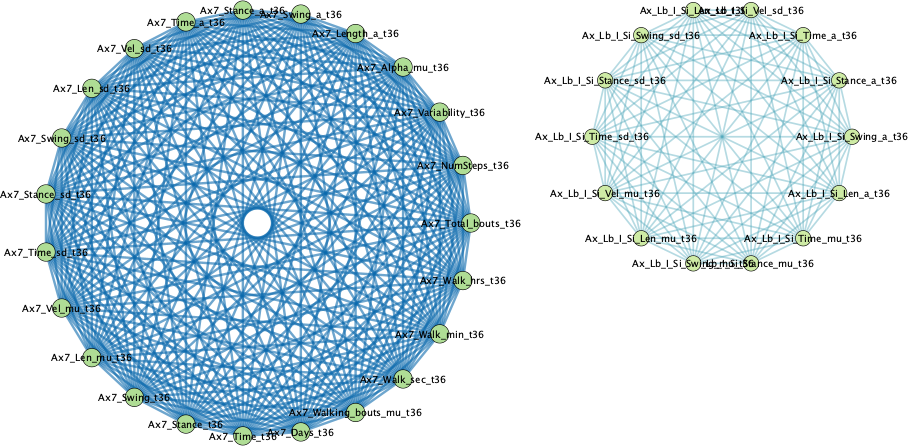}
\caption{Network visualization of the ICICLE dataset with edges filtered by $Q_{JMabs}$ to only display attribute pairs with unexpectedly high or low JM. Edges are coloured by $Q_{JMabs}$ with darker blue indicating higher values.}
\label{Fig:ICICLE_JMabsFilter_Network}
\end{figure}

Further exploration of JM structures using Cytoscape (figure \ref{Fig:ICICLE_JMabsFilter_Network}), where edges have been filtered by $Q_{JMabs}$ to display only the attribute pairs with unexpectedly high or low joint missingness, reveals two clusters of attributes of type $Ax7$ (left cluster) and $Ax$\textunderscore$Lb$ (right cluster) at the 36 month time point. $Ax7$ are 7-day free living measures of activity, as mentioned above, with values missing across all $Ax7$ attributes for those participants that were not called to this part of the study. Correspondingly, $Ax$\textunderscore$Lb$ are activity measures that were measured in the lab which only a subset of participants took part in. Notably, participants taking part in the 7-day activity measuring did not take part in the lab based activity measuring and, consequently, participants with recorded values for $Ax7$ attributes at a certain time point will have missing values across $Ax$\textunderscore$Lb$ attributes for the same time point, and vice versa, which explains the clearly separated clusters of attributes with unexpectedly high JM.

\subsection{Conditional Missingness}
Initial filtering based on $Q_{CMdid}$ revealed a limitation of the CM metrics for datasets with high JM. Considering a variable pair ($\vec{d_j}$, $\vec{d_k}$) with a high number of items with jointly missing values and a very small number of items with missing values in $\vec{d_j}$ but recorded in $\vec{d_k}$. In such case, the distribution in $\vec{d_k}$ of items with missing values in $\vec{d_j}$ but recorded in $\vec{d_k}$ will, unavoidably, be dissimilar to the overall distribution of recorded values in $\vec{d_k}$ due to the small number of items. Similarly, the entropy difference is bound to be high. Such patterns will be picked up as strong conditional missingness by the metrics, but cannot be considered representing a generalisable relationship between missing and recorded values in a variable pair,
due to the sparseness of items with missing values in $\vec{d_j}$ but recorded in $\vec{d_k}$. 

\begin{figure}[t]
\centering
\includegraphics[width=8.5cm]{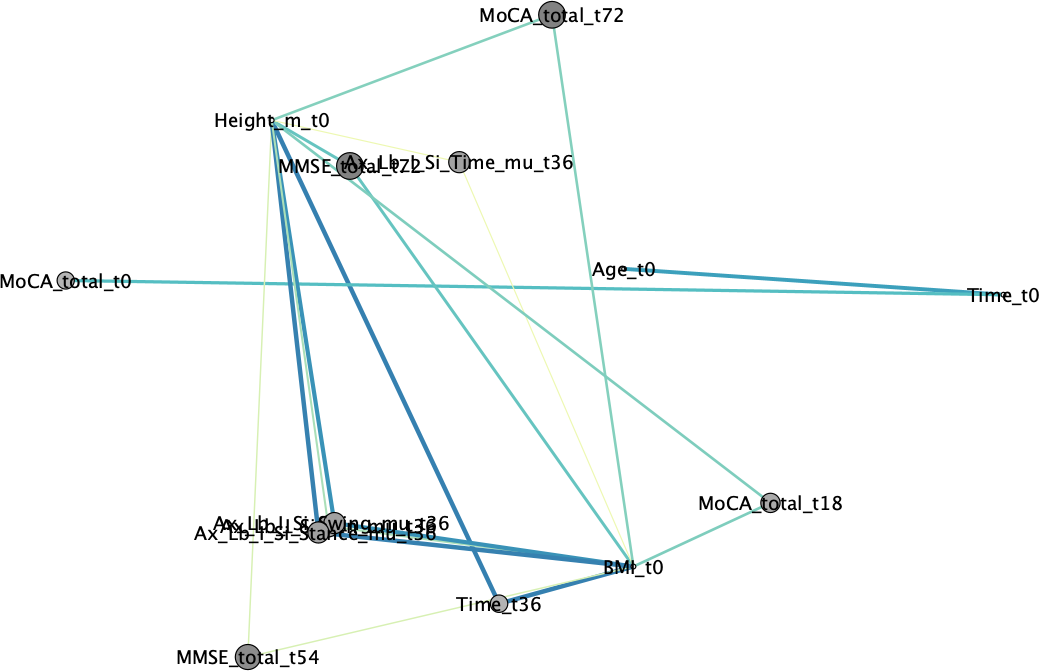}
\caption{Network visualization of the ICICLE dataset with edges filtered by $Q_{JMdir}<0.05$ and $Q_{CMdid}>0.9$, to display attribute pairs with low JM and high CM. Edges are coloured by $Q_{CMdid}$ with dark blue representing higher values.}
\label{Fig:ICICLE_CMdidFilter_Network}
\end{figure}

To mitigate this, an edge filter combination of $Q_{JMdir}<0.05$ and $Q_{CMdid}>0.9$ (i.e., low joint missingness and high conditional missingness) was used in Cytoscape, to enable identification of variable pairs with potentially interesting CM structures. Figure \ref{Fig:ICICLE_CMdidFilter_Network} displays the resulting network, laid out using a force directed layout with edge width and colouring by $Q_{CMdid}$, yellow and green representing lower conditional missingness and dark blue representing high values. The edge colouring in the network reveals a small cluster of nodes towards the bottom left that comprises three $AX\_Lb$ attributes at 36 months with high $Q_{CMdid}$ with the $Height$ and $BMI$ attributes at month 0, as indicated by the connecting dark blue edges. Similarly, the network indicates a strong CM between the $Height$ and $BMI$ attributes at month 0 and $Time$ at month 36.

\begin{figure}[t]
\centering
\includegraphics[width=8.5cm]{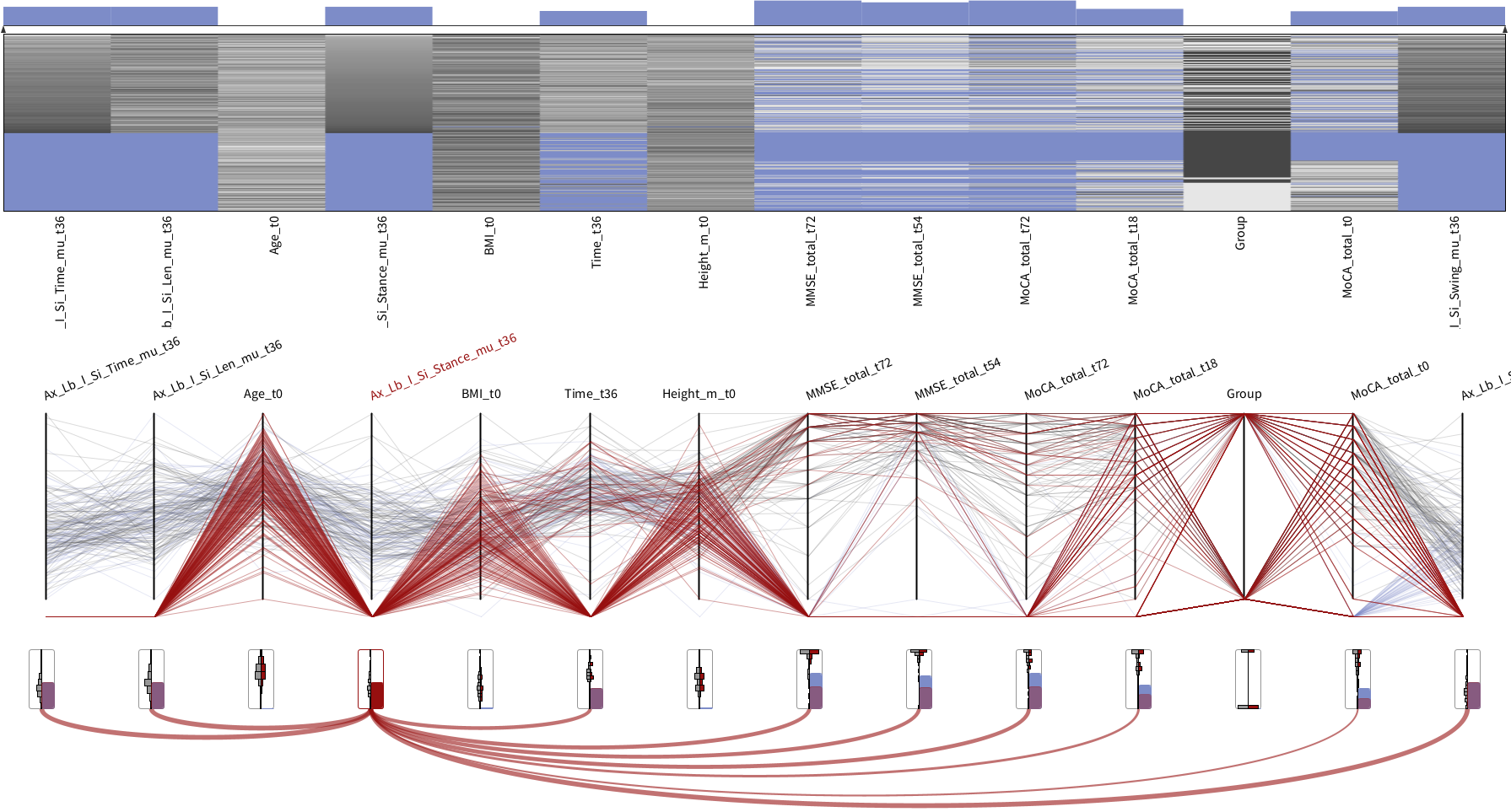}
\caption{Visualization of the subset of variable pairs with $Q_{JMdir}<0.05$ and $Q_{CMdid}>0.9$ displayed in figure \ref{Fig:ICICLE_CMdidFilter_Network}. The fourth attribute from the left is selected, and items and structures relating to missing values in this attribute are highlighted in red in PC and MissiG.}
\label{Fig:ICICLE_CMdidFilter}
\end{figure}

The subset of attribute pairs with low JM and high CM, displayed in figure \ref{Fig:ICICLE_CMdidFilter_Network}, are further investigated using barchart, heatmap, PC and MissiG in figure \ref{Fig:ICICLE_CMdidFilter}. In the figure, the $Time\_t0$ attribute has been removed, as it is the baseline for subsequent gait assessment and is zero for all participants, furhtermore the $Group$ attribute has been added to enable investigation in relation to if participants are healthy controls or have Parkinson's disease. One of the clustered $AX\_Lb$ attributes ($Ax\_Lb\_I\_Si\_Stance\_mu\_t36$, fourth from left) is selected, with structures and items relating to its missing values highlighted in red in the PC and MissiG. Records with missing in this attribute are fairly equally distributed across healthy controls and Parkinson's disease, as visible from the $Group$ attribute (third from right in figure \ref{Fig:ICICLE_CMdidFilter}), indicating that missingness in this attribute is not related to if participants are in the control group or not, differently from e.g. the example in figure \ref{Fig:ICICLE_AMorder_HMPC}.  

\begin{figure}[t]%
       \centering
       \subfloat[][]{
       \includegraphics[height=3cm]{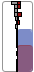}
       	\label{Fig:ICICLE_MoCA_T72}
       }%
       \qquad
        \subfloat[][]{
       \includegraphics[height=3cm]{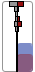}
       	\label{Fig:ICICLE_MoCA_T18}
       }%
       \qquad
        \subfloat[][]{
       \includegraphics[height=3cm]{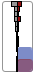}
       	\label{Fig:ICICLE_MoCA_T0}
       }
       \caption{The MissiG representation for $MoCA\_total\_t72$ (\ref{Fig:ICICLE_MoCA_T72}), $MoCA\_total\_t18$ (\ref{Fig:ICICLE_MoCA_T18}) and $MoCA\_total\_t0$ (\ref{Fig:ICICLE_MoCA_T0}) with missing in $Ax\_Lb\_I\_Si\_Stance\_mu\_t36$ highlighted (zoomed in from figure \ref{Fig:ICICLE_CMdidFilter}.
       }
       \label{Fig:ICICLE_MoCA}%
\end{figure}

The distribution of red lines along the $BMI\_t0$ (fifth from left) and $Height\_m\_t0$ (seventh from left) attributes indicate that patients with missing values for this $AX\_Lb$ attribute at month 36 tend to not have high values for $BMI$ and $Height$ at the start of measuring (month 0). Additionally, the MissiG representations for $MoCA$ at month 72, 18 and 0 (fifth, fourth and second axis from the right), which are zoomed in in figure \ref{Fig:ICICLE_MoCA}, reveal a potentially interesting pattern. $MoCA$ (Montreal Cognitive Assessment) is a measure of cognitive efficiency where lower scores indicate more severe cognitive impairment. Comparing the distribution of recorded values in the $MoCA$ attributes for items with missing values in $Ax\_Lb\_I\_Si\_Stance\_mu\_t36$, as represented by the red histograms in the MissiG in figure \ref{Fig:ICICLE_MoCA}, with the distribution of all recorded variables in the $MoCA$ attributes, as represented by the grey histograms in the MissiG, it appears that the red histograms have relatively fewer high values for $MoCA$ attributes at these time points compared to the overall distribution for all participants in the study. 

While no clinical conclusions can be drawn from the findings without further investigation, it can be noted that study participants with missing values for these $Ax\_Lb$ attributes at 36 months appear to have slightly lower BMI and height at the start of the study, as well as slightly lower cognitive efficiency. A hypothesis could be that these participants may have been worse impaired at the start of the study and subsequently more of them had to withdraw from the study due to declining health earlier than participants who were less impaired at the start.

\subsection{Case Study Summary}
This case study provided examples of how the proposed QM can be used to explore and reveal structural missingness in a high-dimensional dataset from a real-life walking monitoring dataset. It demonstrated the ability of the metrics to effectively reveal missingness structures which were confirmed to be related to data collection procedures, but also highlighted limitations of the CM metrics when applied to data with high joint missingness, indicating the need of the analyst to understand the concepts of missingness structures and how the QM may need to be combined to enable identification of interesting and potentially generalisable insights.

\section{Conclusions and Future Work}
\label{sec:conclusions}
This paper presented a set of Quality Metrics that captures aspects of structural missingness in incomplete data, related to the Amount Missing, Joint Missingness and Conditional Missingness structures suggested by Johansson Fernstad \cite{Fernstad2019}. The utilisation and effectiveness of these metrics for visual investigation of structural missingness was demonstrated both for synthetically generated and controlled missingness structures and through a case study with high-dimensional data from a real-life walking monitoring study. The case study contextualized some of the complexities of missing data analysis, and exemplified how QM in general can be applied to a range of visualization aspects, including attribute selection and ordering as well as setting of visual channels such as colour and size, to facilitate identification of potentially interesting structures. 

While the use case presented here provide some evidence of the usefulness of the proposed Quality Metrics, further evaluation and generalisation are needed to fully establish their benefit to missing data visualization. The focus of this would lie on additional exemplar use cases with real life datasets from other domains that display other aspects of missingness structures.

\section*{Supplemental Materials}
\label{sec:supplemental_materials}
All supplemental materials are available at \url{https://doi.org/10.25405/data.ncl.c.7741829}, released under a CC BY 4.0 license.
In particular, they include three CSV files containing the datasets with synthetically generated missingness structures described in section \ref{sec:dataGen}, and a document containing larger versions of the figures in sections \ref{sec:visualinvestigation} and \ref{Sec:ICICLE} as well as additional figures to complement figure \ref{Fig:ICICLE_AMorder_HMPC}.

\section*{Acknowledgments}{%
The authors would like to thank the participants and assessors of the ICICLE and ICICLE-GAIT study. The ICICLE study was supported by Parkinson’s UK (J-0802, G-1301) and by the NIHR Newcastle Biomedical Research Centre at Newcastle Upon Tyne Hospital NHS Foundation Trust and Newcastle University. The work was also supported by the NIHR/Wellcome Trust Clinical Research Facility infrastructure at Newcastle upon Tyne Hospitals NHS Foundation Trust. All opinions are those of the authors and not the funders.%
}

\bibliographystyle{abbrv-doi-hyperref}

\bibliography{refs}

\begin{thebibliography}{10}

\bibitem{alsufyani2024vis}
S.~Alsufyani, M.~Forshaw, S.~Del~Din, A.~Yarnall, L.~Rochester, and S.~J. Fernstad.
\newblock Multi-level visualization for exploration of structures in missing data.
\newblock {\em Computer Graphics and Visual Computing (CGVC). The Eurographics Association}, 2024.

\bibitem{Alsufyani2024star}
S.~Alsufyani, M.~Forshaw, and S.~Johansson~Fernstad.
\newblock Visualization of missing data: a state-of-the-art survey.
\newblock {\em arXiv preprint arXiv:2410.03712}, 2024.

\bibitem{Andreasson2014}
R.~Andreasson and M.~Riveiro.
\newblock Effects of visualizing missing data: an empirical evaluation.
\newblock In {\em 2014 18th International Conference on Information Visualisation}, pp. 132--138. IEEE, 2014.

\bibitem{artero2006enhanced}
A.~O. Artero, M.~C.~F. de~Oliveira, and H.~Levkowitz.
\newblock Enhanced high dimensional data visualization through dimension reduction and attribute arrangement.
\newblock In {\em Tenth International Conference on Information Visualisation (IV'06)}, pp. 707--712. IEEE, 2006.

\bibitem{bauerle2022did}
A.~B{\"a}uerle, C.~van Onzenoodt, S.~der Kinderen, J.~J. Westberg, D.~J{\"o}nsson, and T.~Ropinski.
\newblock Where did my lines go? visualizing missing data in parallel coordinates.
\newblock In {\em Computer Graphics Forum}, vol.~41, pp. 235--246. Wiley Online Library, 2022.

\bibitem{Becker1987}
R.~A. Becker and W.~S. Cleveland.
\newblock Brushing scatterplots.
\newblock {\em Technometrics}, 29(2):127--142, May 1987.

\bibitem{behrisch2018}
M.~Behrisch, M.~Blumenschein, N.~W. Kim, L.~Shao, M.~El-Assady, J.~Fuchs, D.~Seebacher, A.~Diehl, U.~Brandes, H.~Pfister, et~al.
\newblock Quality metrics for information visualization.
\newblock In {\em Computer Graphics Forum}, vol.~37, pp. 625--662. Wiley Online Library, 2018.

\bibitem{Bertini2011}
E.~Bertini, A.~Tatu, and D.~Keim.
\newblock Quality metrics in high-dimensional data visualization: An overview and systematization.
\newblock {\em Visualization and Computer Graphics, IEEE Transactions on}, 17(12):2203--2212, 2011.

\bibitem{Carpenter2013}
J.~Carpenter and M.~Kenward.
\newblock {\em Multiple Imputation and its Application}.
\newblock Wiley, 2013.

\bibitem{Djurcilov2000}
S.~Djurcilov and A.~Pang.
\newblock Visualizing sparse gridded data sets.
\newblock {\em IEEE Computer Graphics and Applications}, 20(5):52--57, 2000.

\bibitem{fernstad2021explore}
S.~J. Fernstad and J.~J. Westberg.
\newblock To explore what isn't there --- glyph-based visualization for analysis of missing values.
\newblock {\em IEEE Transactions on Visualization and Computer Graphics}, 28(10):3513--3529, 2021.

\bibitem{Fielding2009}
S.~Fielding, P.~M. Fayers, and C.~R. Ramsay.
\newblock Investigating the missing data mechanism in quality of life outcomes: a comparison of approaches.
\newblock {\em Health and Quality of Life Outcomes}, 7(1):57, 2009.

\bibitem{Inselberg1985}
A.~Inselberg.
\newblock The plane with parallel coordinates.
\newblock {\em The Visual Computer}, 1(4):69--91, 1985.

\bibitem{jimenez2022graphical}
E.~Jim{\'e}nez and R.~Mac{\'\i}as.
\newblock Graphical tools for visualization of missing data in large longitudinal phenomena.
\newblock In {\em Computer Graphics Forum}, vol.~41, pp. 438--452. Wiley Online Library, 2022.

\bibitem{Johansson2009}
S.~Johansson and J.~Johansson.
\newblock Interactive dimensionality reduction through user-defined combinations of quality metrics.
\newblock {\em IEEE Transactions on Visualization and Computer Graphics}, 15(6):993--1000, 2009.

\bibitem{Fernstad2019}
S.~{Johansson Fernstad}.
\newblock To identify what isn't there: A definition of missingness patterns and evaluation of missing value visualization.
\newblock {\em Information Visualization}, 18(2):230--250, 2019.

\bibitem{Fernstad2014}
S.~{Johansson Fernstad} and R.~C. Glen.
\newblock Visual analysis of missing data -- to see what isn't there.
\newblock In {\em Poster Proceedings of IEEE Vis}. IEEE, November 2014.

\bibitem{Fernstad2011}
S.~{Johansson Fernstad}, J.~Johansson, S.~Adams, J.~Shaw, and D.~Taylor.
\newblock Visual exploration of microbial populations.
\newblock In {\em Proceedings of IEEE Symposium on Biological Data Visualization}, pp. 127--134. IEEE, October 2011.

\bibitem{johansson2020quality}
S.~{Johansson Fernstad}, A.~Macquisten, J.~Berrington, N.~Embleton, and C.~Stewart.
\newblock Quality metrics to guide visual analysis of high dimensional genomics data.
\newblock In {\em EuroVis Workshop On Visual Analytics (EuroVA)}, 2020.

\bibitem{Fernstad2013}
S.~{Johansson Fernstad}, J.~Shaw, and J.~Johansson.
\newblock Quality-based guidance for exploratory dimensionality reduction.
\newblock {\em Information Visualization}, 12(1):44--64, Jan 2013.

\bibitem{Krause2016}
J.~Krause, A.~Dasgupta, J.-D. Fekete, and E.~Bertini.
\newblock Seekaview: An intelligent dimensionality reduction strategy for navigating high-dimensional data spaces.
\newblock In {\em LDAV 2016-IEEE 6th Symposium on Large Data Analysis and Visualization}, 2016.

\bibitem{Kullback1951}
S.~Kullback and R.~A. Leibler.
\newblock On information and sufficiency.
\newblock {\em The Annals of Mathematical Statistics}, 22(1):79--86, 1951.

\bibitem{Lehmann2015}
D.~J. Lehmann, S.~Hundt, and H.~Theisel.
\newblock A study on quality metrics vs. human perception: Can visual measures help us to filter visualizations of interest?
\newblock {\em it-Information Technology}, 57(1):11--21, 2015.

\bibitem{Lewis2012}
J.~M. Lewis, M.~Ackerman, and V.~R. de~Sa.
\newblock Human cluster evaluation and formal quality measures: A comparative study.
\newblock In {\em CogSci}, pp. 1870--1875, 2012.

\bibitem{Liu2017}
S.~Liu, D.~Maljovec, B.~Wang, P.-T. Bremer, and V.~Pascucci.
\newblock Visualizing high-dimensional data: Advances in the past decade.
\newblock {\em IEEE Transactions on Visualization and Computer Graphics}, 23(3):1249--1268, 2017.

\bibitem{lord_cognition_2014}
S.~Lord, B.~Galna, S.~Coleman, A.~Yarnall, D.~Burn, and L.~Rochester.
\newblock Cognition and {Gait} {Show} a {Selective} {Pattern} of {Association} {Dominated} by {Phenotype} in {Incident} {Parkinson}'s {Disease}.
\newblock {\em Frontiers in Aging Neuroscience}, 6, Oct. 2014. \href{https://doi.org/10.3389/fnagi.2014.00249}
{doi: {{%
10\hspace{.1pt}\discretionary{.}{%
}{.}\hspace{.4pt}3389\discretionary{/}{%
}{/}fnagi\hspace{.1pt}\discretionary{.}{%
}{.}\hspace{.4pt}2014\hspace{.1pt}\discretionary{.}{%
}{.}\hspace{.4pt}00249}}}


\bibitem{patricio2018using}
M.~Patr{\'\i}cio, J.~Pereira, J.~Cris{\'o}stomo, P.~Matafome, M.~Gomes, R.~Sei{\c{c}}a, and F.~Caramelo.
\newblock Using resistin, glucose, age and bmi to predict the presence of breast cancer.
\newblock {\em BMC cancer}, 18(1):29, 2018.

\bibitem{Rubin1976}
D.~B. Rubin.
\newblock Inference and missing data.
\newblock {\em Biometrika}, 63(3):581--592, 1976.

\bibitem{Sedlmair2012}
M.~Sedlmair, A.~Tatu, T.~Munzner, and M.~Tory.
\newblock A taxonomy of visual cluster separation factors.
\newblock In {\em Computer Graphics Forum}, vol.~31, pp. 1335--1344. Wiley Online Library, 2012.

\bibitem{Shannon1948}
C.~E. Shannon.
\newblock A mathematical theory of communication.
\newblock {\em The Bell System Technical Journal}, 17(3):379--423, 1948.

\bibitem{shannon2003cytoscape}
P.~Shannon, A.~Markiel, O.~Ozier, N.~S. Baliga, J.~T. Wang, D.~Ramage, N.~Amin, B.~Schwikowski, and T.~Ideker.
\newblock Cytoscape: a software environment for integrated models of biomolecular interaction networks.
\newblock {\em Genome research}, 13(11):2498--2504, 2003.

\bibitem{Shimazaki2007}
H.~Shimazaki and S.~Shinomoto.
\newblock A method for selecting the bin size of a time histogram.
\newblock {\em Neural Computation}, 19(6):1503--1527, 2007.

\bibitem{song2021understanding}
H.~Song, Y.~Fu, B.~Saket, and J.~Stasko.
\newblock Understanding the effects of visualizing missing values on visual data exploration.
\newblock In {\em 2021 IEEE Visualization Conference (VIS)}, pp. 161--165. IEEE, 2021.

\bibitem{song2018}
H.~Song and D.~A. Szafir.
\newblock Where's my data? evaluating visualizations with missing data.
\newblock {\em IEEE transactions on visualization and computer graphics}, 25(1):914--924, 2018.

\bibitem{tierney2023expanding}
N.~Tierney and D.~Cook.
\newblock Expanding tidy data principles to facilitate missing data exploration, visualization and assessment of imputations.
\newblock {\em Journal of Statistical Software}, 105:1--31, 2023.

\bibitem{Turkay2011}
C.~Turkay, P.~Filzmoser, and H.~Hauser.
\newblock Brushing dimensions -- a dual visual analysis model for high-dimensional data.
\newblock {\em IEEE Transactions on Visualization and Computer Graphics}, 17(12):2591--2599, 2011.

\bibitem{Turkay2012}
C.~Turkay, J.~Parulek, and H.~Hauser.
\newblock Dual analysis of dna microarrays.
\newblock In {\em Proceedings of the 12th International Conference on Knowledge Management and Knowledge Technologies}, pp. 26:1--26:8, 2012.

\bibitem{valero2019plot}
P.~Valero-Mora, M.~F. Rodrigo, M.~Sanchez, and J.~SanMartin.
\newblock A plot for the visualization of missing value patterns in multivariate data.
\newblock {\em Practical Assessment, Research, and Evaluation}, 24(1):9, 2019.

\bibitem{Wang2007}
H.~Wang and S.~Wang.
\newblock Visualization of the critical patterns of missing values in classification data.
\newblock In {\em Advances in Visual Information Systems}, pp. 267--274. Springer, 2007.

\bibitem{Wang2009}
H.~Wang and S.~Wang.
\newblock Data mining with incomplete data.
\newblock In {\em Encyclopedia of Data Warehousing and Mining}, pp. 526--530. IGI Global, second ed., 2009.

\bibitem{Wang2019}
J.~Wang, X.~Liu, and H.~W. Shen.
\newblock {High-dimensional data analysis with subspace comparison using matrix visualization}.
\newblock {\em Information Visualization}, 18(1):94--109, 2019. \href{https://doi.org/10.1177/1473871617733996}
{doi: {{%
10\hspace{.1pt}\discretionary{.}{%
}{.}\hspace{.4pt}1177\discretionary{/}{%
}{/}1473871617733996}}}


\bibitem{yarnall_characterizing_2014}
A.~J. Yarnall, D.~P. Breen, G.~W. Duncan, S.~Y. Coleman, C.~Nombela, T.~W. Robbins, K.~Wesnes, D.~J. Brooks, R.~A. Barker, and D.~J. Burn.
\newblock Characterizing mild cognitive impairment in incident {Parkinson} disease: {The} {ICICLE}-{PD} {Study}.
\newblock {\em Neurology}, 82(4):308--316, Jan. 2014. \href{https://doi.org/10.1212/WNL.0000000000000066}
{doi: {{%
10\hspace{.1pt}\discretionary{.}{%
}{.}\hspace{.4pt}1212\discretionary{/}{%
}{/}WNL\hspace{.1pt}\discretionary{.}{%
}{.}\hspace{.4pt}0000000000000066}}}


\end{thebibliography}

\end{document}